\RequirePackage[hyphens]{url}
\documentclass[9pt]{livecoms}

\usepackage[italic]{mathastext}
\usepackage[version=4]{mhchem}
\usepackage{siunitx}
\graphicspath{{figures/}}

\newcommand{\versionnumber}{0.1}  
\newcommand{\githubrepository}{\url{https://github.com/openforcefield/review-protein-benchmark-datasets}}  

\title{Structure-Based Experimental Datasets for Benchmarking Protein Simulation Force Fields [Article v\versionnumber]}

\author[1*]{Chapin E. Cavender}
\author[2]{David A. Case}
\author[3]{Julian C.-H. Chen}
\author[4]{Lillian T. Chong}
\author[5]{Daniel A. Keedy}
\author[6]{Kresten Lindorff-Larsen}
\author[7]{David L. Mobley}
\author[8]{O. H. Samuli Ollila}
\author[9]{Chris Oostenbrink}
\author[10]{Paul Robustelli}
\author[11]{Vincent A. Voelz}
\author[12]{Michael E. Wall}
\author[12]{David C. Wych}
\author[1*]{Michael K. Gilson}

\affil[1]{Skaggs School of Pharmacy and Pharmaceutical Sciences, University of California San Diego, La Jolla, CA, USA}
\affil[2]{Department of Chemistry \& Chemical Biology, Rutgers University, Piscataway, NJ, USA}
\affil[3]{Bioscience Division, Los Alamos National Laboratory, Los Alamos, NM, USA; 
Department of Chemistry and Biochemistry, The University of Toledo, Toledo, OH, USA}
\affil[4]{Department of Chemistry, University of Pittsburgh, Pittsburgh, PA, USA}
\affil[5]{Structural Biology Initiative, CUNY Advanced Science Research Center, New York, NY, USA; 
Department of Chemistry and Biochemistry, City College of New York, New York, NY, USA; 
PhD Programs in Biochemistry, Biology, and Chemistry, CUNY Graduate Center, New York, NY, USA}
\affil[6]{Linderstr\o m-Lang Centre for Protein Science, Department of Biology, University of Copenhagen, Copenhagen N, Denmark}
\affil[7]{Department of Pharmaceutical Sciences, University of California Irvine, Irvine, CA, USA}
\affil[8]{Institute of Biotechnology, University of Helsinki, Helsinki, Finland; VTT Technical Research Centre of Finland, Espoo, Finland}
\affil[9]{Institute for Molecular Modeling and Simulation, University of Natural Resources and Life Sciences, Vienna, Austria}
\affil[10]{Department of Chemistry, Dartmouth College, Hanover, NH, USA}
\affil[11]{Department of Chemistry, Temple University, Philadelphia, PA, USA}
\affil[12]{Computer, Computational, and Statistical Sciences Division, Los Alamos National Laboratory, Los Alamos, NM, USA; 
The Center for Nonlinear Studies, Los Alamos National Laboratory, Los Alamos, NM, USA}

\corr{chapin.cavender@openforcefield.org}{CEC}
\corr{mgilson@health.ucsd.edu}{MKG}

\orcid{Chapin E. Cavender}{0000-0002-5899-7953}
\orcid{David A. Case}{0000-0003-2314-2346}
\orcid{Julian C.-H. Chen}{0000-0003-0341-165X}
\orcid{Lillian T. Chong}{0000-0002-0590-483X}
\orcid{Daniel A. Keedy}{0000-0002-9184-7586}
\orcid{Kresten Lindorff-Larsen}{0000-0002-4750-6039}
\orcid{David L. Mobley}{0000-0002-1083-5533}
\orcid{O.H. Samuli Ollila}{0000-0002-8728-1006}
\orcid{Chris Oostenbrink}{0000-0002-4232-2556}
\orcid{Paul Robustelli}{0000-0002-9282-8993}
\orcid{Vincent A. Voelz}{0000-0002-1054-2124}
\orcid{Michael E. Wall}{0000-0003-1000-688X}
\orcid{David C. Wych}{0000-0001-9209-4371}
\orcid{Michael K. Gilson}{0000-0002-3375-1738}

\pubDOI{10.XXXX/YYYYYYY}
\pubvolume{<volume>}
\pubissue{<issue>}
\pubyear{<year>}
\articlenum{<number>}
\datereceived{Day Month Year}
\dateaccepted{Day Month Year}

\begin{document}

\begin{frontmatter}
\maketitle

\begin{abstract}
This review article provides an overview of structurally oriented experimental datasets that can be used to benchmark protein force fields, focusing on data generated by nuclear magnetic resonance (NMR) spectroscopy and room temperature (RT) protein crystallography.
We discuss what the observables are, what they tell us about structure and dynamics, what makes them useful for assessing force field accuracy, and how they can be connected to molecular dynamics simulations carried out using the force field one wishes to benchmark.
We also touch on statistical issues that arise when comparing simulations with experiment.
We hope this article will be particularly useful to computational researchers and trainees who develop, benchmark, or use protein force fields for molecular simulations.
\end{abstract}

\end{frontmatter}

\clearpage

\tableofcontents
\begin{quote}
{\em It is a truth universally acknowledged that a research group in possession of a good force field must be in want of a benchmark.}
\end{quote}

\section{Introduction}

The earliest computer simulations of biomolecules could explore only sub-nanosecond phenomena \cite{mccammon_dynamics_1977,case_dynamics_1979,levitt_accurate_1988}, but advances in computing hardware \cite{stone_accelerating_2007,shaw_anton_2008} and simulation algorithms \cite{darden_particle_1993,wennberg_direct-space_2015} now enable the simulation of processes that occur on biologically relevant timescales, enabling quantitative studies of, for example, protein conformational changes \cite{grant_large_2010,anandakrishnan_speed_2015}, ligand binding \cite{gilson_calculation_2007,wang_identifying_2013,reif_net_2014}, protein folding \cite{lindorff-larsen_how_2011,shaw_atomic-level_2010}, and the assembly of multi-protein complexes \cite{ssaglam_proteinprotein_2019}.
As a consequence, simulations have become increasingly useful tools to improve our understanding of protein functions, elucidate molecular mechanisms of human disease, and design small molecule drugs that work by binding targeted proteins.
Although knowledge-based heuristic approaches\cite{rohl_protein_2004} and approaches based on deep learning \cite{jumper_highly_2021} may be the current methods of choice for predicting the most stable conformation of a protein \cite{kryshtafovych_critical_2021}, studying the thermodynamics of phenomena such as conformational changes and ligand binding still requires simulation-based sampling of conformations away from the global energy minimum.

Molecular simulations, such as molecular dynamics and Monte Carlo simulations, are, in effect, importance-weighted sampling methods \cite{frenkel_understanding_2001}.
They sample conformations from a protein’s Boltzmann-weighted ensemble, where the Boltzmann weight is based on a potential function, also known as an energy model.
These conformations, or snapshots, make up a simulation trajectory, and can be used to compute physical observables as Boltzmann averages of quantities estimated from the individual snapshots, because most experimental observables are averages over time for many molecules. The more conformations are sampled, the greater the numerical precision of the physical property estimates.
Given a sufficiently large number of conformations, the reliability of the predictions is no longer determined by amount of sampling, and depends instead on the accuracy of the energy model and the method used to compute the property of interest from the trajectory.
Not surprisingly, there is a strong tradeoff between precision, which is determined by the amount of sampling done, and accuracy, because more accurate energy models are more computationally costly and thus reduce the amount of sampling that can be achieved.
For example, quantum chemical energy models can be highly accurate but remain too slow to provide adequately converged estimates of many properties of interest or to simulate phenomena on biologically relevant time-scales.
Therefore, the simulation community has a continued interest in far more computationally efficient energy models called force fields.
These are parameterized, physics-based energy models that use simple approximations of interatomic interactions and thus can be evaluated quickly, enabling relatively fast conformational sampling.

Force fields specifically for proteins composed of the 20 canonical amino acids date back to the 1980s and 1990s, and their parameters were typically derived from quantum chemical calculations or from the bulk properties of neat liquids chosen as small molecule analogs of protein fragments \cite{jorgensen_opls_1988,cornell_second_1995,mackerell_all-atom_1998}.
The quality of these force fields was then evaluated by running simulations of proteins in water and computing the root mean square deviation (RMSD) of the simulated protein’s atomic coordinates from the corresponding coordinates of a static crystal structure of the protein.
Over time, a growing collection of structural and dynamical data on proteins, generated by X-ray diffraction and nuclear magnetic resonance (NMR) experiments, has enabled more detailed evaluations of the conformational ensembles produced by protein force fields.
Deficiencies identified in these assessments have motivated further tuning of protein force fields, such as the addition of protein-specific corrections to the force fields’ torsional parameters \cite{lindorff-larsen_improved_2010,best_optimization_2012,mackerell_jr_extending_2004,Hornak:2006:Proteins,Maier:2015:J.Chem.TheoryComput.,diem_hamiltonian_2020,tian_ff19sb_2020}.
Further information on the history of today’s protein force fields is available in prior reviews \cite{lifson_recent_1973,ponder_force_2003,guvench_comparison_2008,zhu_recent_2012,dauber-osguthorpe_biomolecular_2019}.

Although modern protein force fields have been fruitfully applied, they still provide only approximations to the true quantum mechanical energies, and it is not possible to predict {\em a priori} the consequences of the approximations made for the calculation of quantities of interest. 
Additionally, the different research groups that develop force fields often prioritize conflicting goals, such as accuracy for the simulation of folded versus disordered proteins or for the prediction of structure versus thermodynamic properties.
A given protein force field therefore models some observables better and others less well.
Thus, to obtain a well-rounded picture of a force field’s strengths and weaknesses, it is essential to benchmark it against a variety of data types. 

In this article, we review and describe experimental data that interrogate a range of structural and dynamical features of proteins and therefore are useful to benchmark the accuracy of protein force fields.
We focus on experimental observables that provide detailed information about protein conformational ensembles under conditions similar to those of greatest interest for protein simulations, i.e., well-hydrated proteins near room temperature and pressure.
We consider only datasets that involve water-soluble (i.e. non-membrane bound) proteins and peptides without ligands or cofactors, because when non-protein molecules (e.g. membrane lipids or drug molecules) are present in the simulation, the results depend on not only the protein force field but also the force field used for these other compounds.
That said, since we are interested in proteins in water, the accuracy of the simulations necessarily also depends on the choice of water model.
This article is intended for computational researchers who develop, benchmark, or use protein force fields.
We thus assume familiarity with molecular dynamics techniques, force field terms, and the basics of protein structure, but not with the experimental techniques used to generate the data.

The remainder of the review is organized into four sections.
Section \ref{sec:overall} summarizes key points and recommendations.
Section \ref{sec:nmr} describes experiments using nuclear magnetic resonance (NMR) spectroscopy, and Section \ref{sec:xtal} describes experiments using room-temperature (RT) crystallography.
In each of these two sections, we first review the types of observables provided by the experiments, why they are useful for interrogating protein conformational dynamics, and how they can be calculated from simulation trajectories.
Then, we describe specific datasets containing measurements of these observables for peptide or protein systems, as summarized in Table \ref{tab:datasets}.
Finally, Section \ref{sec:best_practices} discusses best practices for setting up and analyzing benchmark simulations.

\section{Overall Recommendations for Choosing Benchmark Datasets}
\label{sec:overall}
 
Although early benchmarks of protein force fields compared simulations of solvated proteins to corresponding X-ray structures---i.e., to single structural models from crystal diffraction experiments---more recent benchmarks utilize experimental observables from solution experiments like NMR.
Such experiments match the conditions of dilute aqueous solutions most commonly used for protein simulations, and NMR observables include information about excursions from native structures that contribute to solution-phase ensembles. 
In contrast, there are fewer studies that evaluate the ability of force fields to reproduce diffraction observables in crystal simulations.
Crystal simulations are not applicable to disordered proteins and require larger and longer simulations to converge estimates of observables. In addition, methods of comparing experimental observables with those estimated from simulations are not as well developed.
Nonetheless, we have included a section on crystal observables because these provide additional information on the structure, interactions, and dynamics of proteins.
As computational throughput continues to increase we hope that benchmarks targeting crystal observables become more commonplace.

\begin{table*}[!ht]
\centering
\begin{tabular}{p{0.3 \textwidth} p{0.3 \textwidth} p{0.3 \textwidth}}
\toprule
{\bf Dataset} & {\bf Observables} & \bf{Description} \\
\midrule
\multicolumn{3}{c}{\bf Nuclear magnetic resonance spectroscopy} \\
\midrule
Beauchamp short peptides \cite{beauchamp_are_2012} & Chemical shifts, J-couplings & Short, unstructured peptides \\
Designed $\beta$-hairpins \cite{blanco_short_1994,ramirez-alvarado_novo_1996,de_alba_turn_1997,maynard_origin_1998,stanger_rules_1998,cochran_tryptophan_2001,ramirez-alvarado_elongation_2001,pastor_combinatorial_2002,russell_stability_2003,skelton_hairpin_2003,fesinmeyer_enhanced_2004,espinosa_autonomously_2005,andersen_minimization_2006,honda_crystal_2008,kier_stabilizing_2010,scian_mutational_2013,jimenez_design_2014,anderson_nascent_2016,morales_design_2019,richaud_folding_2021,peintner_pushing_2022} and Trp-cage miniproteins \cite{struthers_design_1996,mcknight_nmr_1997,struthers_design_1998,imperiali_uniquely_1999,vermeulen_solution_2004,cornilescu_solution_2007,gronwald_evolutionary_2008,neidigh_designing_2002,lin_helical_2004,barua_trp-cage_2008,scian_crystal_2012,byrne_circular_2013,graham_reversing_2019} & Chemical shifts, NOEs, folding stabilities & Small, folded peptides with mutations \\
Stroet folded proteins \cite{stroet_validation_2024} & J-couplings, RDCs, NOEs & Folded proteins \\
Mao folded proteins \cite{mao_protein_2014} & Chemical shifts, NOEs & Diverse folded proteins \\
Robustelli a99SB-disp \cite{robustelli_developing_2018} & Chemical shifts, J-couplings, RDCs, NOEs, PREs & Folded proteins and disordered proteins \\
Spin relaxation datasets \cite{jarymowycz_fast_2006,louhivuori2006conformational,ciragan_nmr_2020,khan_distribution_2015} & Spin relaxation rates & Folded proteins and disordered proteins \\
Salt bridge stabilities \cite{tomlinson2009characterization} & Salt bridge stabilities derived from chemical shifts & Folded proteins \\
\midrule
\multicolumn{3}{c}{\bf Room-temperature crystallography} \\
\midrule
Scorpion toxin \cite{smith_ab_1997} & Electron density, B-factors & Folded protein with little secondary structure \\
Hen egg white lysozyme \cite{ramanadham_refinement_1990,meisburger_diffuse_2020,walsh_refinement_1998,wang_triclinic_2007,artymiuk_structures_1982,meisburger_robust_2023} & Electron density, B-factors, diffuse scattering & Rigid, folded protein \\
Crambin \cite{teeter_water_1984,chen_room-temperature_2012} & Electron density, B-factors, neutron scattering & Rigid, folded protein \\
Cyclophilin A \cite{keedy_mapping_2015,van_benschoten_measuring_2016,thompson_temperature-jump_2019,chen_solvent_2024} & Electron density, B-factors, alternate conformations & Folded protein with multiple accessible states and crystal data at different temperatures \\
Ubiquitin \cite{biel_flexibility_2017} & Electron density, B-factors, alternate conformations & Folded protein with mutations \\
PTP1B \cite{keedy_expanded_2018} & Electron density, B-factors, alternate conformations & Folded protein with multiple accessible states and crystal data at different temperatures \\
Endoglucanase \cite{nakamura_newtons_2015} & Electron density, B-factors, neutron scattering & Folded protein with measured solvent density \\
Staphylococcal nuclease \cite{wall_conformational_2014} & Electron density, B-factors, diffuse scattering & Folded protein \\
\bottomrule
\end{tabular}
\caption{Summary of experimental datasets for benchmarking protein force fields.}
\label{tab:datasets}
\end{table*}

Table \ref{tab:datasets} summarizes many existing experimental datasets---both NMR and 
crystallographic---suitable for benchmark studies of protein force fields.
We recommend that a minimal benchmark include at least one dataset containing peptides, folded proteins, and disordered proteins---because a generally applicable protein FF should work for all three cases.
We further recommend using datasets that target the following observables, as listed in the second column of Table \ref{tab:datasets}:

\begin{itemize}
\item chemical shifts
\item scalar couplings
\item residual dipolar couplings
\item electron and nuclear densities from Bragg diffraction
\end{itemize}

\begin{table*}[!ht]
\centering
\begin{tabular}{p{0.35 \textwidth} c c c c}
\toprule
{\bf Observable} & {\bf Experimental} & \bf Convergence & \bf Structural & \bf Consensus on \\
& \bf uncertainty & & \bf information & \bf best practices \\
\midrule
\multicolumn{5}{c}{\bf Nuclear magnetic resonance spectroscopy} \\
\midrule
Chemical shifts & Low & Fast & Local & Yes \\
Scalar couplings & Low & Fast & Local & Yes \\
Residual dipolar couplings & Low & Fast & Tertiary & Yes \\
Nuclear Overhauser effect spectroscopy & High & Slow & Tertiary & No \\
Spin relaxation & High & Slow & Tertiary & No \\
Paramagnetic relaxation enhancement & High & Slow & Tertiary & No \\
\midrule
\multicolumn{5}{c}{\bf Room-temperature crystallography} \\
\midrule
Bragg diffraction (X-ray and neutron) & Low & Slow & Tertiary & No \\
B-factors & High & Fast & Local & Yes \\
Alternate conformations & Low & Slow & Local & No \\
Diffuse scattering & High & Slow & Tertiary & No \\
\bottomrule
\end{tabular}
\caption{Summary of features of experimental observables.}
\label{tab:observables}
\end{table*}

The first three are useful because they have low experimental uncertainty, fast convergence in simulations, and a consensus on best practices for estimating the observables.
Although Bragg diffraction lacks such a consensus and converges slowly, it stands out from the other observables considered in this review because it informs on protein tertiary structures while having low experimental uncertainty.
See Table \ref{tab:observables} for a summary of the salient qualitative features of each observable: whether the experimental uncertainty is high or low, whether the estimate of the observable from simulations converges quickly or slowly, whether the observable primarily provides information about the local structure or tertiary structure of proteins, and whether there is a consensus in the literature on best practices for comparing simulations to experiments.

The other observables (nuclear Overhauser effect intensities, spin relaxation rates, paramagnetic relaxation enhancements, crystallographic B-factors, crystallographic alternate conformations, diffuse crystallographic scattering intensities) are less ideal because each exhibits some combination of high experimental uncertainties, slow convergence in simulations, or lack of consensus on best practices for estimating the observable.
We therefore recommend that they be deprioritized under circumstances of limited computational resources.
Nonetheless, these observables can provide useful alternative characterizations of protein structural ensembles.
Note, too, that some can be estimated from the same trajectories used to estimate observables in the first list.
For example, simulations used to estimate Bragg diffraction intensities could also be used to estimate B factors.
One approach is to perform a two-tier benchmark, in which the first tier uses the first list of observables to assess many candidate force fields, and the second tier uses the second list to assess force fields that perform well on the first tier.

Finally, although one may aspire to compare molecular simulations "directly" with experiment, this is impossible.
Experimental observables are raw data like radio frequency signals collected during NMR studies and speckle patterns collected during crystallographic experiments, while simulations give atomic coordinates over time.
Thus, models, with their own assumptions and parameters, are needed to bring calculation together with experiment for comparison.
For example, computing chemical shifts from molecular simulations requires a model (Section \ref{sub2:chem_shift}), and a protein crystal structure is a model based on diffraction data (Section \ref{sub2:bragg}).
Thus, the agreement between experimental and computed observables should be interpreted with an understanding of the models used in estimating the observable from a simulation and the experimental observable itself from the raw data, as previously emphasized \cite{van_gunsteren_deriving_2016,van_gunsteren_validation_2018}.

\section{Nuclear magnetic resonance (NMR) spectroscopy}
\label{sec:nmr}

Nuclear magnetic resonance (NMR) spectroscopy measures the responses of nuclear magnetic moments in a strong external magnetic field to perturbations by weak oscillating external magnetic fields tuned to the resonant frequency of the nuclei.
The observed responses are sensitive to the local magnetic fields at the nuclei, which in turn depend on what other nuclei are nearby.
As a consequence, NMR spectroscopy provides information about the local chemical environments of atoms and hence about the conformational distributions and dynamics of the molecules they belong to.

NMR is applicable to nuclei with an odd number of nucleons and hence nonzero nuclear spin---such as $^1$H, $^{13}$C, $^{15}N$, $^{19}$F, and $^{31}$P---because these possess a magnetic dipole moment.
In the presence of a strong external magnetic field, this dipole precesses around the external field, much as a spinning top precesses around the downward gravitational field.
The angular frequency of the precession, which is called the Larmor frequency, is proportional to the strength of the magnetic field at the nucleus, with a proportionality constant characteristic of the nuclear isotope.
If a second, weak magnetic field that oscillates near the Larmor frequency is applied, the axis of precession will rotate away from the direction of the strong external field.
An NMR spectrometer detects the resulting transverse magnetization by measuring the electrical current it induces in a coil of wire.
Although the value of the Larmor frequency is determined chiefly by the strong external magnetic field, the electronic structure near each nucleus, which is influenced by through-bond and through-space interactions, modulates the field felt by the nucleus and hence its Larmor frequency, so NMR probes the local environment of the nucleus.
Additionally, magnetization can be transferred to nearby nuclei via through-space interactions, so NMR can report on the relative dispositions in space of pairs of nuclei.
These two effects give NMR spectroscopy its sensitivity to the structure and dynamics of proteins.
In practice, specific sequences of radiofrequency pulses are crafted to interrogate various aspects of these relaxation phenomena. 

NMR observables have several useful features that have led to their adoption as targets for both training and validation of protein force fields \cite{allison_assessing_2012}.
First, NMR experiments are typically performed in laboratory conditions that are similar to the desired setup for most simulation applications, namely dilute aqueous solution.
Note, however, that NMR studies are often done at moderately low pH (often around pH 6), to minimize effects from amide proton exchange with solvent, so it is important that simulations meant for comparison against NMR data assign pH-appropriate protonation states of titratable residues and uncapped protein termini.
Second, in contrast with typical X-ray or neutron crystallography experiments, NMR spectroscopy can provide useful information about disordered proteins, i.e., proteins that do not fold into a well-defined structure which can crystallize.
Furthermore, whereas other methods applicable to disordered proteins, such as small angle X-ray scattering, provide only low-resolution structural information, NMR observables can report on structural features that are closely related to specific force field terms, such as the probability distributions of rotatable bonds, which connect closely with the torsional energy terms for a particular dihedral angle.
Finally, because NMR observables are averages over ensembles that include deviations from native structures, they can report on the thermodynamic balance among native states, near-native states with local rearrangements, and unfolded states, a perspective not readily available from crystallography.
Indeed, early protein force fields tended to maintain the correct structures of folded proteins but had trouble correctly sampling structural fluctuations.
Thus, they often overestimated local fluctuations of folded proteins about their mean structures yet overestimated the amount of structure in  disordered regions of proteins \cite{mackerell_jr_extending_2004,Hornak:2006:Proteins,lindorff-larsen_improved_2010,best_optimization_2012,Maier:2015:J.Chem.TheoryComput.,diem_hamiltonian_2020}.
Such deficiencies were identified and largely corrected by comparisons with NMR observables that report on the fluctuations in folded proteins and the amount of residual structure in more disordered regions.

Here, Section \ref{sub:nmr_obs} discusses NMR observables that can be used to assess the accuracy of molecular simulations and hence of the force fields used in simulations.
The observables considered are chemical shifts, scalar couplings (also known as J-couplings), residual dipolar couplings (RDCs), the nuclear Overhauser effect (NOE), spin relaxation, and paramagnetic relaxation enhancement (PRE).
Section \ref{sub:nmr_data} then presents available experimental NMR datasets that are well suited for evaluating simulations.

\subsection{NMR observables}
\label{sub:nmr_obs}

\subsubsection{Chemical shifts}
\label{sub2:chem_shift}

\paragraph{General principles}

The chemical shift of a nucleus---the difference of its Larmor frequency from that of the same isotope in a reference compound---probes the degree to which the nuclear spin "feels" the externally imposed magnetic field.
This is determined by its electronic environment, which in turn is controlled by the details of the local molecular structure.
In a protein, one can consider the chemical shift of a nucleus as having a baseline offset resulting from its local covalent connectivity and bond hybridization, and an additional shift determined by both its local geometry (bond lengths, bond angles, and dihedral angles) and through-space interactions resulting from electric fields, hydrogen bonds, and the proximity of chemical groups that contain substantial magnetic anisotropy, such as aromatic rings \cite{wishart_12_1994,wishart_1_2002,neal_rapid_2003}.
The sensitivity of a given nucleus to each of these influences depends on its chemical identity (atomic number), covalent structure (bond hybridization), and chemical environment.
If $\omega$ is the observed Larmor frequency of a nucleus in a particular molecular context, and the reference Larmor frequency for the same isotope is $\omega_0$, the chemical shift, $\delta$, is reported as

\begin{equation}
\label{eqn:chem_shift}
\delta = \frac {\omega - \omega_0} {\omega_0}
\end{equation}

\noindent Because $\omega$ and $\omega_0$ are similar, $\delta$ is typically reported in parts per million (ppm).

\begin{figure}[t]
    \centering
    \includegraphics[width=3 in]{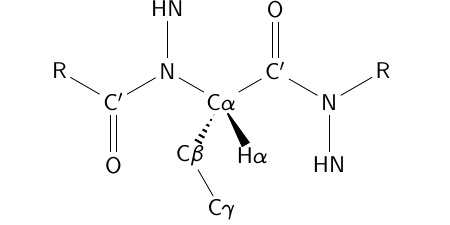}
    \caption{Atom names and named dihedral angles associated with commonly measured chemical shifts and $^3J$-couplings in a typical amino acid.
    Sidechain hydrogens such as $H\beta$ are omitted for clarity.}
    \label{fig:dihedrals}
\end{figure}

Empirical algorithms have been developed to predict the chemical shifts of protein backbone atoms for a given set of three-dimensional coordinates.
These algorithms---implemented in software packages that include SHIFTS \cite{xu_automated_2001}, PROSHIFT \cite{meiler_proshift_2003}, Camshift \cite{kohlhoff_fast_2009}, SPARTA+ \cite{shen_sparta_2010}, SHIFTX2/SHIFTX+ \cite{han_shiftx2_2011}, PPM/PPM\_One \cite{li_ppm_2012,li_ppm_one_2015}, UCB-Shift \cite{li_accurate_2020}, and GraphNMR \cite{yang_predicting_2021}---are trained on databases of proteins for which both high resolution X-ray structures and solution backbone NMR chemical shift assignments are available.
They take protein coordinates as inputs, and output a chemical shift prediction for the backbone nuclei of C$\alpha$, C$\beta$, C$'$, N, HN, and H$\alpha$ atoms (Figure\ref{fig:dihedrals}).
Each nucleus type has a baseline offset, $\delta_{random}$, determined by its identity, that reflects the chemical shift expected if the protein were in a random coil conformation, and this baseline offset is modified by additive terms for specific structural features.
The structural features with the largest influence on the backbone nuclei of a given residue are the backbone and sidechain dihedral angles of that residue and of neighboring residues within two positions in primary sequence; the distances and orientations of nearby aromatic rings and other chemical groups with substantial magnetic anisotropy; the presence and geometry of hydrogen bonds; the proximity of polar and charged nuclei; and the solvent exposure of the residue.

Of these, the main determinant of backbone chemical shifts are local dihedral angles (see Figure\ref{fig:dihedrals}).
The C$\alpha$ and C$\beta$ shifts of residue $i$ are most sensitive to the $\phi$ and $\psi$ angles of the same residue, $\phi_i$ and $\psi_i$.
C$'$ shifts are most sensitive to the $\psi$ angles of residue $i$ and the following residue $i+1$, $\psi_i$ and $\psi_{i+1}$.
N shifts are most sensitive to the $\chi_1$ angle of that residue and the $\psi$ angle of the preceding residue $i-1$, $\chi_{1i}$ and $\psi_{i-1}$.
Proton shifts are more sensitive to non-bonded interactions than are those of carbon and nitrogen atoms.
In particular, HN and H$\alpha$ shifts are very sensitive to the presence of aromatic ring currents, HN shifts are particularly sensitive to hydrogen bond geometries, and H$\alpha$ shifts are sensitive to electric fields from nearby polar and charged atoms.

The field of empirical protein backbone chemical shift prediction is relatively mature, with a high degree of consensus in the predictions of algorithms published in the last 20 years.
Three of the first broadly applicable protein chemical shift prediction algorithms are SHIFTS \cite{xu_automated_2001}, SHIFTX \cite{neal_rapid_2003}, and PROSHIFT \cite{meiler_proshift_2003}. Of these, SHIFTX generally produces the most accurate predictions on protein structures not contained in its training databases, and it requires only a few seconds to predict all backbone shifts in a protein.
Later, the SPARTA algorithm \cite{shen_protein_2007} provided a small improvement in prediction accuracy, with somewhat slower calculation times.
The program Camshift \cite{kohlhoff_fast_2009}, published in 2009, produced comparable accuracy to SHIFTX and SPARTA, and utilizes interatomic distance-based equations that can be evaluated in milliseconds and are differentiable with respect to atomic coordinates, enabling the computationally efficient incorporation of chemical shifts as structural restraints in MD simulations.
Taken together, SPARTA, SHIFTX, and Camshift represented important milestones in the field, as their predictions were good enough to enable the prediction of accurate protein structures using only NMR chemical shifts combined with molecular mechanics force fields or knowledge-based potential energy functions \cite{cavalli_protein_2007,shen_consistent_2008,wishart_cs23d_2008,robustelli_folding_2009,robustelli_using_2010}.
Additionally, Camshift expresses chemical shifts as polynomials of interatomic distances, and so Camshift predictions are differentiable with respect to atomic coordinates and can be used as restraints in dynamics simulations.
SPARTA, SHIFTX, and Camshift were also found to be sensitive to the conformational fluctuations of proteins observed in MD simulations, and they were utilized to computationally generate protein conformational ensembles that accurately model the dynamics of proteins \cite{li_certification_2010,markwick_enhanced_2010,robustelli_interpreting_2012,robustelli_conformational_2013} and to guide the optimization of protein force field torsion terms \cite{li_nmr-based_2010,robustelli_developing_2018}.
PPM was parameterized using MD simulations as input (rather than X-ray structures), thus aiming to link conformational ensembles to chemical shifts \cite{li_ppm_2012}.

A more recent generation of empirical shift predictors---including SPARTA+ \cite{shen_sparta_2010}, SHIFTX2/SHIFTX+ \cite{han_shiftx2_2011}, PPM\_One \cite{li_ppm_one_2015}, UCB-Shift \cite{li_accurate_2020,ptaszek_ucbshift_2024}, and GraphNMR \cite{yang_predicting_2021}---were developed using machine learning techniques (although the earlier method PROSHIFT \cite{meiler_proshift_2003} is also based on a neural network).
They provide improved accuracy, have very similar accuracy to one another, and typically produce very similar results in practical applications such as the validation of MD ensembles, the fitting of force field corrections, the calculation of protein structures and structural ensembles, and the reweighting of MD trajectories.
Despite the accuracy of these methods, their errors are still an order of magnitude larger than the experimental uncertainties in the chemical shifts themselves, so experimental uncertainties can generally be neglected when comparing calculation to experiment.

In addition to empirical prediction algorithms trained on databases of protein structures, other prediction algorithms use quantum mechanical calculations on local fragments of proteins to estimate chemical shifts.
Quantum predictors aim to estimate the shielding of the magnetic moment of a nucleus due to its local environment in a particular conformation, and these methods are sensitive to small changes in the coordinates of the input structure.
While quantum predictors can achieve good agreement with experimental chemical shifts when averaged over many conformations in a protein's ensemble \cite{yi_contribution_2024,yi_predicted_2024}, they typically provide lower accuracy and computational speed than empirical predictors \cite{mulder_nmr_2010,frank_toward_2012,sumowski_sensitivity_2014,fritz_determination_2018,case_using_2020,chandy_accurate_2020}.

Programs have also been developed to calculate the NMR chemical shifts of protein side-chain atoms, but they tend to be less informative than backbone chemical shift predictors \cite{han_shiftx2_2011,sahakyan_structure-based_2011,sahakyan_using_2011,li_ppm_one_2015,ptaszek_ucbshift_2024}.
This is because the predictors tend to be less accurate, due to a smaller database of training data, and because the experimental chemical shifts for each atom type tend to have small ranges and can take on similar values for different side chain conformations.

\paragraph{Evaluation of MD simulations using chemical shifts}

The chemical shift of a given nucleus in a protein can be estimated from a simulation as a simple average of the predicted chemical shifts over the $N_{\mathsf{frame}}$ simulation frames: 

\begin{equation}
\label{eqn:shift_estimate}
\langle \delta \rangle = \frac {1} {N_\mathsf{frame}} \sum_{t=1}^{N_{\mathsf{frame}}} \delta_t
\end{equation}

\noindent Here $t$ indexes simulation frames and $\delta_t$ is the chemical shift predicted for simulation frame $t$ using one of the methods discussed above.
Deviations between chemical shifts predicted from MD trajectories and experimental shifts are sensitive indicators of errors in MD ensembles.
Accordingly, NMR chemical shift predictions are routinely used to assess the accuracy of MD simulations, and torsion terms in protein force fields have been parameterized by optimizing the agreement between experiment and simulation \cite{li_nmr-based_2010,robustelli_developing_2018}.

Backbone chemical shift predictions computed from MD ensembles that have large deviations from experiment---i.e. greater than one- to two-fold higher than the average predictor accuracy for a given backbone atom type (e.g. C$\alpha$) as assessed on a predictor’s training databases---indicate substantial inaccuracies in the simulated ensemble.
The largest deviations generally reflect incorrect secondary structures and/or side-chain rotamers.
Smaller deviations can result from inaccurate distributions of tertiary interactions in the MD ensemble, including hydrogen bonds, electrostatic interactions from charged groups, and aromatic ring currents.
The latter are particularly important for proton chemical shifts.
For disordered proteins, backbone chemical shifts are very sensitive to the presence of partially populated secondary structure elements, and deviations from experiment can indicate that MD ensembles overestimate or underestimate rare $\alpha$-helical, PP$_\mathsf{II}$, or $\beta$-sheet propensities.

When comparing calculated and experimental backbone chemical shifts, one should keep the following considerations in mind.

First, chemical shift prediction algorithms are imperfect, so isolated errors in individual atoms or residues may not imply problems with a simulation.
However, a series of residues with several atoms having large deviations between predicted and experimental shifts probably does reflect an error in the conformational distribution of the simulation.
The average prediction accuracy on the database of X-ray structures used to train a given predictor provides a reasonable baseline to identify problematic simulations.
For example, the standard deviation of C$\alpha$ shift predictions made by SPARTA+ is 0.94 ppm for its training database, so a contiguous stretch of residues with C$\alpha$ prediction errors greater than 1.5 ppm probably indicates a problem with the simulated ensemble, whereas a simulation where the errors are less than 1.00 ppm is relatively reliable.

Second, as noted above, the chemical shifts of the backbone nuclei (atom types C$\alpha$, C$\beta$, C$'$, N, HN, H$\alpha$) of each residue type have large baseline offsets, $\delta_{\mathsf{random}}$, which are determined by chemical identity, i.e. by their covalent structure rather than by any time-dependent conformational variation that might be sampled in a simulation \cite{schwarzinger_sequence-dependent_2001,de_simone_accurate_2009,tamiola_sequence-specific_2010,kjaergaard_sequence_2011}.
The random coil shifts of a given backbone atom type can vary more strongly across residue types (e.g. ~50 ppm for C$\beta$ between Ala and Thr \cite{kjaergaard_disordered_2012}) than across residues of a given type (e.g. 3 ppm to 5 ppm for Ala C$\beta$ atoms) due to conformational variations \cite{ulrich_biomagresbank_2008,romero_biomagresbank_2020}.
As a result, it can be highly misleading to visualize correlations and report correlation coefficients between experimental chemical shifts and shifts predicted from MD simulations, as most of the magnitude in deviation between different residues can be explained by differences in their baseline $\delta_{\mathsf{random}}$ values.
Indeed, extremely high correlation coefficients (close to 1.0) can be obtained by comparing experimental chemical shifts to database random coil chemical shift values without utilizing any structural information \cite{de_simone_accurate_2009,kohlhoff_fast_2009,shen_sparta_2010,han_shiftx2_2011}.
A more informative visualization of the accuracy of chemical shift predictions from an MD simulation is provided by comparing the deviations between predicted and experimental shifts separately for each residue type.
Alternatively, one can compare predicted and experimental “secondary” chemical shifts, which subtract the random coil chemical shift values from both experimental and predicted shifts and thus put all residues on comparable footing.

Third, empirical chemical shift predictions depend on many structural features, so one cannot be sure what conformational error causes a given prediction error. For example, a deviation of 1.5 ppm for an N atom may result from a side chain populating an incorrect rotamer or an aromatic group being incorrectly positioned.

Fourth, chemical shift prediction error is sequence- and conformation-specific and so should not be used to compare the accuracy of simulations of two different proteins or of two different regions of one protein.
For example, chemical shift prediction errors for $\beta$-sheet proteins are substantially higher than for $\alpha$-helical proteins \cite{yang_predicting_2021}, so a simulation of a $\beta$-sheet protein may yield worse agreement with experimental chemical shifts than an equally accurate simulation of an $\alpha$-helical protein.
In contrast, when one protein is simulated with multiple force fields, the accuracy of the chemical shift predictions is a clear indication of the relative accuracy of the simulations and hence of the force fields.
Note that scalar couplings do not have this limitation, as detailed in Section \ref{sub2:j_coupling}.

Fifth, in principle, chemical shift predictions computed from accurate, thermalized conformational ensembles should be more accurate than predictions computed from static protein structures.
However, the most accurate chemical shift prediction algorithms are trained against static X-ray structures (with PPM being one exception \cite{li_ppm_2012}).
As a result, some amount of conformational averaging is "baked in" to these prediction algorithms; that is, chemical shifts predicted from static conformations already implicitly account for the solution-phase averaging over conformations.
This fact may help account for the fact that chemical shifts predicted based on X-ray structures are often (though not always) more accurate than predictions from MD ensembles \cite{li_certification_2010,markwick_enhanced_2010,robustelli_interpreting_2012,robustelli_conformational_2013}.
However, the lower accuracy obtained with simulations does not necessarily point to inaccuracies of simulations.
Indeed, MD simulations that yield excellent agreement with other NMR observables (such as NMR scalar couplings or residual dipolar couplings; see below), and that are thus presumably quite accurate, may produce less accurate chemical shift predictions than static high-resolution X-ray structures \cite{li_certification_2010,robustelli_interpreting_2012,robustelli_developing_2018}.

Sixth, comparisons between simulated and experimental chemical shifts are particularly informative regarding the accuracy of simulations of disordered proteins and peptides.
The lack of stable tertiary structure in such systems means that through-space interactions are largely washed out so that chemical shifts are dominated by backbone dihedral angles, which are the most accurately parametrized and least noisy relationships in backbone chemical shift predictors \cite{salmon_nmr_2010}.
As a result, chemical shifts computed for disordered proteins can be more accurate than chemical shifts computed for training databases of X-ray structures of folded proteins.
For example, C$'$ chemical shift predictions obtained from simulations of disordered proteins and peptides that agree with orthogonal NMR data or with circular dichroism data frequently have RMSDs from experiment less than 0.5 ppm, whereas C$'$ prediction errors for folded proteins average ~1 ppm.
Indeed, simulations of disordered proteins with state-of-the art force fields regularly achieve prediction RMSDs a factor of 2 lower than the average predictor errors observed on databases of folded proteins, with a substantial dynamic range such that agreement with chemical shifts correlates well with the agreement with orthogonal experimental data \cite{salmon_nmr_2010,robustelli_conformational_2013,schwalbe_predictive_2014,robustelli_developing_2018}.
Therefore, when several different force fields yield predicted chemical shifts for disordered proteins with RMSDs lower than typically observed for folded proteins, differences across the force fields being tested should not be dismissed as being within prediction error but instead should be regarded as meaningful and informative.

Seventh, there is not a one-to-one mapping between conformational distributions and chemical shifts \cite{ozenne2012mapping}. As a consequence, many different conformational ensembles can produce equivalent agreement with experiment \cite{wood_secondary_2011,virtanen_heterogeneous_2020,shrestha_full_2021}.
For example, the same C$'$ chemical shift prediction can be obtained from many different $\phi$/$\psi$ dihedral distributions; and two ensembles with extremely different $\phi$/$\psi$ dihedral distributions for a given residue can produce identical chemical shift predictions for a C$\alpha$ atom if, for example, it is exposed to an aromatic ring current in one ensemble and not the other.
The degeneracy between conformations and chemical shifts can often be resolved by analyzing the chemical shifts of multiple nuclei.

Eighth, different programs for predicting chemical shifts tend to yield very similar results, so the choice of method is not expected to have much effect on conclusions.
However, some chemical shift predictors utilize sequence homology for part or all of their predictions, and a prediction based on sequence homology rather than three-dimensional structure \emph{ipso facto} does not report on the accuracy of a structure.
For example, the SHIFTX2 predictor is a weighted average between a structure-based predictor SHIFTX+ and a homology-based predictor SHIFTY.
It is recommended to use only structure-based predictors for validation of MD simulations and to use the same method consistently across the trajectories being compared.

\paragraph{Evaluation of force fields via structural properties derived from chemical shifts}
\label{sub3:derived-chemical-shifts}

Measured chemical shifts can be used to derive structural information, such as the helicity of a peptide in solution.
Simulations can then be evaluated based on their agreement with these experimentally derived structural data.
This indirect approach avoids the complications of predicting chemical shifts from simulations, instead relying on the availability of reliable methods to map from chemical shifts to structures.
Two examples of structural models that can be used in this way, helical propensities and stabilities of salt bridges and salt bridge analogs, are now considered.

Accurate protein force fields should be able to model the preferences for proteins to adopt particular secondary structures.
Helical propensity in a particular sequence context is often modeled using the expected fraction of time that a particular non-terminal residue adopts an $\alpha$-helical backbone conformation, and the fractional helicity of a residue can be measured because $\alpha$-helical residues form backbone hydrogen bonds that alter the chemical shift of the C$'$ carbons in $^{13}$C-labeled proteins \cite{shalongo_distribution_1994}.
We note here that this commonly used dataset was measured in D$_2$O as a solvent.
Since the helicity will likely be different in H$_2$O and D$_2$O \cite{shi2002d}, one should not expect perfect agreement in simulations with water models that represent H$_2$O.
Assuming a two-state helix-coil transition, the helical fraction can be calculated using

\begin{equation}
\label{eqn:fraction_helix}
f_{\mathsf{helix}} = \frac {\delta_{\mathsf{obs}} - \delta_{\mathsf{coil}}} {\delta_{\mathsf{helix}} - \delta_{\mathsf{coil}}}
\end{equation}

\noindent where $\delta_{\mathsf{obs}}$,  $\delta_{\mathsf{helix}}$, and $\delta_{\mathsf{coil}}$, are, respectively, the C$'$ chemical shifts observed in the experiment, in the reference helical state, and in the reference coil state.
Meanwhile, a simulation can be analyzed to provide the fraction of time each residue is in a helical conformation, based on its backbone dihedral angles or hydrogen bond occupancies, and the results can be compared with the results inferred from chemical shifts \cite{best_optimized_2009}.
Alternatively, researchers can fit simulated conformations to a helix-coil transition model such as the Lifson-Roig model \cite{lifson_theory_1961}.
A commonly used variant of the Lifson-Roig model expresses the partition function of a two-state helix-coil system in terms of two parameters: a nucleation parameter describing the likelihood for a residue to sample the helical dihedral angles in the coil state and an extension parameter describing the likelihood for a neighboring residue at the end of a helix to transition from the coil to the helix state.
These two parameters can be allowed to take different values for each residue type, and the resulting set of parameters can be fit to the fraction of helix from the simulation and then compared to parameters derived from experimental data \cite{best_optimized_2009}.

Another important characteristic of protein force fields is the ability to accurately model the formation of salt bridges---pairs of amino acids whose oppositely charged side-chains are within hydrogen bonding distance \cite{donald_salt_2011}.
When a salt bridge is formed, the presence of the anionic side chain alters the chemical shift of nitrogen in the cationic side chain, and this perturbation can be measured in $^{15}$N-labeled proteins \cite{tomlinson2009characterization} or in small molecule analogs of these side chains.
Similar to the helical fraction, the fraction of salt bridge formation can be calculated from the chemical shifts observed in the experiment, along with reference shifts measured in the presence and absence of the salt bridge.
Simulations can then be analyzed to provide the fraction of time the salt bridge is present based on geometric criteria for hydrogen bonding between the charged side chains.

\subsubsection{Scalar couplings}
\label{sub2:j_coupling}

\paragraph{General principles}

NMR scalar couplings, also known as indirect couplings or $J$-couplings, are electron-mediated, spin-spin couplings which act particularly strongly through covalent bonds and hence are often described as through-bond couplings \cite{karplus_contact_1959,karplus_vicinal_1963}. Scalar couples occur when the magnetic field of a nuclear spin modifies the behavior of nearby electrons and this modification propagates to nearby nuclei, resulting in an indirect nucleus-nucleus coupling.
Scalar couplings can be observed in an NMR spectrum as the splitting of the resonance peak of one nucleus into two (sometimes overlapping) peaks.
The value of the scalar coupling, also called the coupling constant, is the distance between the split peaks in the frequency domain and thus has units of reciprocal time (e.g. Hz).
These couplings can propagate through multiple bonds, and three-bond couplings are of particular interest because they provide a readout of the dihedral angle of the central bond. Accordingly, three-bond couplings, termed $^3J$-coupling constants, have been utilized extensively in the conformational analysis of small molecules \cite{karplus_vicinal_1963} and peptides \cite{bystrov_spinspin_1976} since the discovery of the Karplus relationship \cite{karplus_contact_1959}, or Karplus equation, which relates the $^3J$-coupling constant between two nuclear spins to the intervening dihedral angle $\theta$:

\begin{equation}
\label{eqn:karplus}
^3J(\theta) = A \cos^2 (\theta) + B \cos (\theta) + C
\end{equation}

\noindent Here the Karplus coefficients---$A$, $B$, and $C$---depend on the identities of the nuclei involved in the coupling  (e.g., $^{13}$C, $^1$H) and their local chemical environments, including bond hybridizations, bond lengths, bond angles, and the electronegativities of nearby substituents \cite{haasnoot_relationship_1980}.
There are also one-bond $^1J$ and two-bond $^2J$ scalar couplings, which can have more complex Karplus relationships that depend on more than one dihedral angle \cite{vuister_use_1993,cornilescu_large_2000,wirmer_angular_2002,ding_protein_2004,gapsys_improved_2015}.

Karplus equation coefficients used in conformational analyses are generally empirically determined, based on measurements of $J$-coupling constants in molecules of known structure, and then transferred to analyze the dihedral angles between nuclei in similar chemical environments \cite{karplus_contact_1959,karplus_vicinal_1963,haasnoot_relationship_1980,wang_determination_1996,vogeli_limits_2007}.
For a given pair of coupled nuclei, the experimentally observed $J$-coupling is the probability-weighted average of their instantaneous coupling over the molecule’s thermalized conformational distribution.

In proteins and peptides, the $^3J$-couplings between backbone amide protons and alpha protons ($^3J_{\mathsf{HN-H\alpha}}$) report on the $\phi$ dihedral angle of the peptide backbone (Fig.\ref{fig:dihedrals}) and thus distinguish between $\alpha$ and $\beta$ secondary structure.
These couplings were adopted as structural restraints in early NMR protein structure calculations \cite{pardi_calibration_1984}.
The values of the Karplus coefficients for these $^3J$-coupling constants have been the subject of frequent reexamination and scrutiny \cite{case_static_2000,lindorff-larsen_interpreting_2005,altona_vicinal_2007,vogeli_limits_2007,wang_quantum_2013,lee_quantitative_2015,li_high_2015}, including studies that examine the consequences of harmonic motion and conformational dynamics \cite{brueschweiler_adding_1994,case_static_2000,lindorff-larsen_interpreting_2005,vogeli_limits_2007,lee_quantitative_2015}.
Although $^3J_{\mathsf{HN-H\alpha}}$ are the most frequently measured and reported $^3J$-couplings for the protein backbone, five additional coupling constants also report on the $\phi$ dihedral angle \cite{schmidt_self-consistent_1999}: $^3J_{\mathsf{HN-C'}}$, $^3J_{\mathsf{HN-C\beta}}$, $^3J_{\mathsf{C'(i-1)-H\alpha}}$, $^3J_{\mathsf{C'(i-1)-C'}}$, and $^3J_{\mathsf{C'(i-1)-C\beta}}$.
It has also been shown that one-bond $^1J$-scalar couplings, such as $^1J_{\mathsf{C\alpha-H\alpha}}$ and $^1J_{\mathsf{C\alpha-C\beta}}$, are sensitive to the $\psi$ angle of protein backbones \cite{vuister_use_1993,cornilescu_large_2000,gapsys_improved_2015}.
The $\chi_1$ angles of protein sidechains can be analyzed via the following $^3J$-couplings \cite{perez_self-consistent_2001,chou_insights_2003}: $^3J_{\mathsf{H\alpha-H\beta}}$, $^3J_{\mathsf{N-H\beta}}$, $^3J_{\mathsf{C'-H\beta}}$, $^3J_{\mathsf{H\alpha-C\gamma}}$, $^3J_{\mathsf{N-C\gamma}}$, and $^3J_{\mathsf{C'-C\gamma}}$.
Additionally, $^3J_{\mathsf{HN-C'}}$ hydrogen bond scalar couplings (scalar couplings between protein backbone nitrogen and carbonyl atoms in different residues that are mediated through hydrogen bonds) provide quantitative information about hydrogen bond geometries \cite{barfield_structural_2002}.
Once Karplus coefficients are known, no specialized software is required to compute scalar couplings for a protein structure.
One simply needs to calculate the dihedral angles of interest and use them to predict the corresponding scalar couplings via the Karplus relationship applied with the appropriate Karplus coefficients.
Special attention, however, is needed to ensure correct mapping to the correct atoms in cases when stereospecific assignments are available so that, for example, the same definition of the atoms are used in experiments and simulations.

\paragraph{Evaluation of force fields using scalar couplings}

Observed $J$-coupling constants can be estimated from the snapshots of an MD trajectory as

\begin{equation}
\label{eqn:j_coupling_estimate}
\langle J(\theta) \rangle = \frac {1} {N_\mathsf{frame}} \sum_{t=1}^{N_{\mathsf{frame}}} J(\theta_t)
\end{equation}

\noindent where $t$ indexes the $N_{\mathsf{frame}}$ simulation frames and $J(\theta)$ is given by a Karplus relationship (Eq. \ref{eqn:karplus}).
This mean can be compared with the corresponding measured $J$-coupling.
One may therefore use $J$-coupling data to test \cite{graf_structure_2007,best_are_2008,lindorff2012systematic,beauchamp_are_2012,Maier:2015:J.Chem.TheoryComput.} and parameterize \cite{best_optimized_2009,piana_how_2011,best_optimization_2012,best_balanced_2014,Maier:2015:J.Chem.TheoryComput.,robustelli_developing_2018} simulation force fields.
In such calculations, one must choose between "static" and "dynamics-corrected" Karplus coefficients.
Static coefficients are obtained from empirical fits of ensemble-averaged solution data to high-resolution X-ray structures and therefore do not explicitly account for the complexities of conformational distributions during parametrization, analogous to the "baked in" dynamics of many chemical shift predictors (Section \ref{sub2:chem_shift}).
Dynamics-corrected Karplus coefficients are obtained from empirical fits that seek to account explicitly for the distributions of dihedral angles of the molecule in solution using a variety of approaches, including single- and multiple-well harmonic motion models as well as fitting of coefficients to rotamer population distributions obtained from other types of experimental data \cite{brueschweiler_adding_1994,case_static_2000,chou_insights_2003,lindorff-larsen_interpreting_2005,vogeli_limits_2007,lee_quantitative_2015}.

A recent study \cite{robustelli_developing_2018} examined the accuracy of $^3J$-couplings computed with long MD simulations using several sets of Karplus coefficients and seven different force fields, for a large set of NMR data spanning folded and disordered proteins. The static $^3J_{\mathsf{HN-H\alpha}}$ Karplus coefficients from Vogeli et al \cite{vogeli_limits_2007} and the static $^3J_{\mathsf{C'(i-1)-C'}}$ Karplus coefficients from Li et al \cite{li_high_2015} produced the lowest RMSD values from experiment on average across force fields and protein systems.
The dynamics-corrected Karplus coefficients from Lee et al \cite{lee_quantitative_2015} gave similar trends in accuracy across force fields, but larger average deviations from the experimental couplings.
Karplus coefficients could be derived not empirically but from quantum calculations for specific conformations, thus avoiding the question of conformational averaging.
Karplus coefficients derived in this way have similar values as dynamics-corrected empirical Karplus coefficients \cite{brueschweiler_adding_1994,case_static_2000,chou_insights_2003,lindorff-larsen_interpreting_2005,vogeli_limits_2007,lee_quantitative_2015}.

When interpreting the NMR scalar couplings calculated from MD simulations, it is essential to consider the uncertainties in the Karplus coefficients.
This is frequently done using a $\chi^2$ value \cite{best_balanced_2014,robustelli_developing_2018}:

\begin{equation}
\label{eqn:chi_sq}
\chi^2 = \frac {1} {N_{\mathsf{obs}}} \sum_{i=1}^{N_{\mathsf{obs}}} \frac {\left( \langle J_i(\theta) \rangle - J_{i,\mathsf{exp}} \right)^2} {\sigma_i^2}
\end{equation}

\noindent where $i$ indexes the $N_{\mathsf{obs}}$ observables, $\langle J_i(\theta) \rangle$ and $J_{i,\mathsf{exp}}$ are the computed (Eq. \ref{eqn:j_coupling_estimate}) and experimental $J$-coupling constants, respectively, and $\sigma_i$ is the RMSD between predicted and measured scalar couplings obtained in fitting of the Karplus parameters. A $\chi^2$ value less than 1.0 indicates agreement within the estimated uncertainty. It is also worth recalling the words of Martin Karplus \cite{karplus_vicinal_1963}: "the person who attempts to estimate dihedral angles to an accuracy of one or two degrees does so at \ldots [their] own peril".

\subsubsection{Residual dipolar couplings}
\label{sub2:rdc}

\paragraph{General principles}

The direct interaction of the magnetic dipoles associated with two nuclear spins leads to an experimentally measurable coupling, known as a dipolar coupling.
Similar to scalar couplings, the dipolar coupling can be observed as the splitting of a peak in the frequency domain and thus has units of reciprocal time (Hz).
The magnitude of the dipolar coupling between the nuclear spins of atoms $A$ and $B$ for an instantaneous configuration and orientation of the molecule is given by

\begin{equation}
\label{eqn:rdc}
D_{AB} = D_{AB,\mathsf{max}} \left( 3 \cos^2 \theta_{AB} - 1 \right) = D_{AB,\mathsf{max}} r_{AB}^T \mathcal{A} r_{AB}
\end{equation}

\noindent where $D_{AB,\mathsf{max}}$ depends on the identities of the nuclei and the distance between them, and $\theta_{AB}$ is the angle between the magnetic field imposed by the NMR instrument and the vector connecting atoms A and B \cite{bax2001dipolar,bax2003weak,chiliveri2021advances}.
The second equality expresses the quantity in parentheses in terms of $r_{AB}$, the unit vector joining the two nuclei, defined in the internal coordinates of the molecule, and the alignment tensor $\mathcal{A}$, which relates an internal coordinate system to the lab-frame magnetic field.
In a solution of freely tumbling molecules, $D_{AB}$ takes on all possible values with equal probability, and the observed dipolar couplings---which are averages over molecules and time---are zero.

However, if the molecules can be even weakly aligned relative to the instrument’s magnetic field, then $D_{AB}$ no longer averages to zero, and dipolar couplings can report on the structure and conformational dynamics of proteins.
Such alignment may be achieved by linking the protein to a prosthetic group that tends to align with the field; by placing it in an aqueous liquid crystal formed by, for example, bicelles \cite{sanders_magnetically_1990,prosser_magnetically_1998}; or by anisotropic compression of acrylamide gels \cite{tycko_alignment_2000}.
The dipolar coupling measured in a weakly aligned sample is called a residual dipolar coupling (RDC).
RDCs are often measured between atoms that are directly bonded.
In proteins, these are often an amide proton and nitrogen, resulting in a $^1D_{\mathsf{NH}}$ RDC, but it is also possible to measure RDCs between nuclei that are not directly bonded.
Although the alignment procedure could in principle perturb the protein’s conformational ensemble, consistency across different methods for alignment suggests this is not a substantial concern \cite{lakomek2008self}.

\paragraph{Evaluation of MD simulations using residual dipolar couplings}

To use protein RDCs to benchmark a simulation, one must estimate the alignment tensor, $\mathcal{A}$.
For folded proteins, it is often reasonable to assume that the internal motions of the protein and the alignment are mostly decoupled or that any coupling does not contribute substantially to the RDCs \cite{louhivuori2006conformational,salvatella2008influence}.
In this case, one may keep the concept of an alignment tensor, noting that this should be fitted over the full ensemble rather than using a single structure \cite{lindorff2005simultaneous,showalter2007quantitative}.
The procedure and equations used in this fitting are the same as for rigid proteins and typically rely on alignment of the MD frames, followed by singular value decomposition (SVD) of the unit vectors of the bonds of interest.
With a set of experimental RDCs and a simulation, one can fit the five independent parameters of the alignment tensor.
In contrast, for unfolded proteins, flexible peptides, and intrinsically disordered proteins, it is not possible to fit an average alignment from the data because the alignment varies across conformations.
Therefore, most analyses use a physical model to predict the alignment tensor for each conformation (rather than to fit it from the experiments) and use this to calculate per-frame RDCs that can then be averaged.
These physical models can be tested against data for folded proteins, where the alignment can be determined from experiment using a SVD fitting procedure.
Such tests suggest that the predicted alignments are accurate enough to be useful, in particular when alignment is dominated by steric interactions between the macromolecule and the alignment medium.
Among the different methods for predicting alignment tensors, the PALES software is probably the most commonly used \cite{zweckstetter_nmr_2008}.
While it is possible to predict alignment for the full chain, sometimes local alignment over short stretches of ca. 15 residues is used instead \cite{marsh2008calculation}.

RDCs have been used extensively to benchmark protein force fields \cite{lange2010scrutinizing,lindorff-larsen_improved_2010,lindorff2012systematic,robustelli_developing_2018}, as they offer a number of advantages.
In particular, they can be measured very precisely \cite{chiliveri2021advances}, they are averaged over long time scales and thus report on conformational ensembles \cite{lakomek2008self}, and they can be measured for multiple bonds in both the backbone and side chains \cite{chiliveri2021advances}.

Because the magnitudes of the RDCs depend on the alignment strength \cite{tjandra_direct_1997,trigo-mourino_structural_2011}, the quality of agreement between experiments and simulations is typically evaluated in terms of a normalized RMSD (the quality factor, $Q$ \cite{cornilescu_validation_1998}), with high-resolution crystal structures typically having $Q<25$\% \cite{bax2003weak}.

\subsubsection{Nuclear Overhauser effect spectroscopy}
\label{sub2:noesy}

\paragraph{General principles}

The nuclear Overhauser effect (NOE) is the change in intensity of the resonance peak of one nucleus that occurs when the resonance of a nearby nucleus is saturated by radio frequency irradiation.
Saturation of a nucleus greatly perturbs its magnetization away from its equilibrium value, and the interaction between the magnetic dipoles of two nuclei nearby in space allows the system to relax toward equilibrium via a concerted flip of both nuclear spins.
This concerted flip is known as dipolar cross relaxation and requires an exchange of energy between the two-spin system and nearby atoms.
In NMR spectroscopy, the environment surrounding a system of spins is called the lattice.
(However, this term does not imply a repeating arrangement of atoms related by translational symmetry as in a crystal lattice.)
Thus, dipolar cross relaxation involves coupling to the motions of the lattice. The NOE is a through-space, rather than through-bond, effect that is typically detectable only when the mean internuclear distance is less than \qty{0.6}{\nano\meter}.

\begin{figure}[t]
    \centering
    \includegraphics[width=3 in]{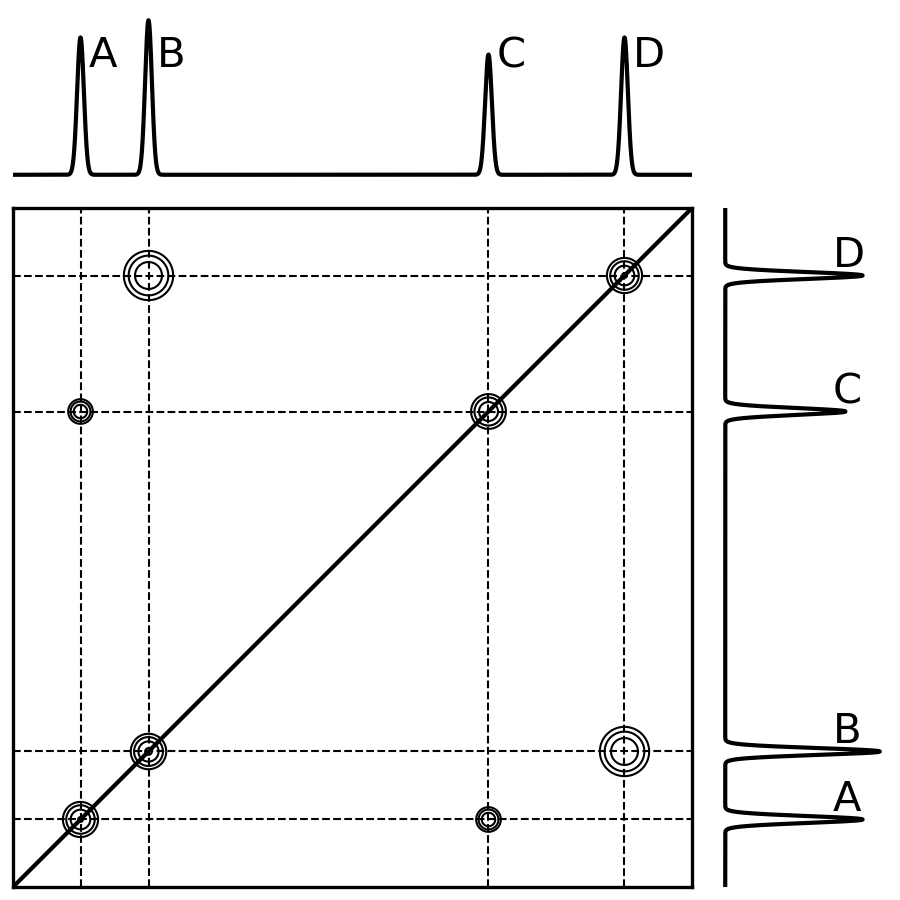}
    \caption{Idealized NOESY spectrum.
    The one-dimensional spectrum displayed along the top and right contains peaks for four hydrogen nuclei, labeled "A" through "D".
    Apart from the self-peaks along the diagonal, there are NOE cross peaks between two pairs of nuclei.
    The volumes of the cross peaks indicate a weak NOE between A and C and a strong NOE between B and D nuclei.
    The lack of cross peaks between A and D and between B and C suggests no significant NOE between these pairs of nuclei.}
    \label{fig:noesy-2d}
\end{figure}

NOEs for proteins are typically measured using a two-dimensional nuclear Overhauser effect spectroscopy (NOESY) experiment, in which one dimension corresponds to the chemical shift of the saturated nucleus, $\delta_A$, and the second dimension corresponds to the chemical shift of the observed nucleus, $\delta_B$.
The intensity of the NOE between nuclei $A$ and $B$ is measured as the integral of the cross peak located at ($\delta_A,\delta_B$) in the two-dimensional NOESY spectrum (Fig. \ref{fig:noesy-2d}).
This integral is proportional to the time elapsed between the saturation and observation pulses, i.e. the mixing time.

NOESY experiments for proteins report on the average, or effective, distances between pairs of nuclei, where the averages are taken over the the mixing time of the NOESY experiment, and the form of the average is determined by the time scales of intramolecular vs overall motions of the protein \cite{neuhaus_nuclear_2000,vogeli_nuclear_2014}.
In the limit where intramolecular motions are much slower than molecular tumbling, the effective distance between nuclei A and B, $R_{AB}$, takes the form

\begin{equation}
\label{eqn:noe_R_slow}
R_{AB}^{\mathsf{slow}} = \left \langle \frac {1} {R_{AB}^6} \right \rangle^{-1/6}
\end{equation}

\noindent In the limit where variations in the orientations of internuclear displacements are small and where intramolecular motions are much faster than molecular tumbling, the average becomes 

\begin{equation}
\label{eqn:noe_R_fast}
R_{AB}^{\mathsf{fast}} = \left \langle \frac {1} {R_{AB}^3} \right \rangle^{-1/3}
\end{equation}

\noindent Appendix \ref{app:noesy_distances} provides additional information about the precise meanings of these assumptions and how they give rise to the corresponding effective distances, and more detailed discussions of this topic are available in the literature \cite{neuhaus_nuclear_2000,vogeli_nuclear_2014}.

The previous discussion has assumed that the cross peak intensities in a NOESY experiment result purely from direct dipolar cross relaxation between the saturated and observed spins.
However, observed NOESY intensities also include contributions from indirect dipolar cross relaxation events mediated by additional spins nearby in space.
For example, after saturation of spin $A$, the cross-peak intensity between spins $A$ and $C$ will include not only a direct contribution but also an indirect contribution due to dipolar cross relaxation between spins $A$ and $B$ followed by dipolar cross relaxation between spins $B$ and $C$.
This indirect relaxation, called spin diffusion, can reduce the internuclear distances inferred from NOEs. 
Although spin diffusion is non-negligible in most protein systems, it is slower than direct relaxation, so it contributes less to NOESY spectra with shorter mixing times, making such experiments more straightforward to interpret and simpler to model with MD simulations.

\paragraph{Evaluation of MD simulations using NOESY}

The literature describes many approaches for assessing the accuracy of a protein ensemble against a NOESY experiment.
Most can be grouped into three broad categories, according to the level of detail used in modeling the dependence of the NOEs on intramolecular motions, as now discussed.

The first approach treats the dependence on intramolecular motions implicitly by assigning the experimental NOESY cross peak intensities to categories associated with upper boundaries on the corresponding effective internuclear distances; e.g., \qty{0.25}{\nano\meter} for strong, \qty{0.30}{\nano\meter} for medium, and \qty{0.45}{\nano\meter} for weak NOEs \cite{smith_structure_1993}.
Appropriate ensemble-averaged effective distances, i.e. $R_{AB}^{\mathsf{slow}}$ (Eq. \ref{eqn:noe_R_slow}) or $R_{AB}^{\mathsf{fast}}$ (Eq. \ref{eqn:noe_R_fast}), can be computed from a simulation and compared to these upper boundaries, and force fields can be assessed by counting violations of these boundaries.
The choice of which distance average to use depends on assumptions about the time scales of intramolecular motions relative to overall molecular tumbling.
For typical globular proteins, time constants for molecular tumbling are on the order of \qtyrange{10}{100}{\nano\second}, and intramolecular motions are assumed to occur on faster time scales so that $R_{AB}^{\mathsf{fast}}$ is an appropriate choice of effective distance.
For short, unstructured peptides, time constants for molecular tumbling are much faster, and intramolecular motions are assumed to occur on slower time scales so that $R_{AB}^{\mathsf{slow}}$ is an appropriate choice of effective distance.
For intrinsically disordered proteins and for folded proteins that undergo slow, large-scale conformational fluctuations, it is likely that neither assumption is valid, and a different approach for evaluating NOESY intensities should be used (see below).
Note that this first approach ignores the effects of spin diffusion and therefore is most appropriate for comparison against NOESY spectra collected at short mixing times.
Sometimes experimental measurements are reported only as upper boundaries on distances rather than as direct NOESY intensities, in which case this approach is the only one available.
The experimental uncertainty in this approach is mostly due to the discretization of the measured intensities into rough distance restraints, which should be kept in mind when interpreting the number and size of any violations between simulated and experimental effective distances.

In the second approach, effective distances are calculated from the experimental NOESY cross peak intensities by assuming that the cross peak intensities $V_{AB}$ are proportional to the reciprocal sixth power of the effective distance.

\begin{equation}
\label{eqn:noesy_sixth_power}
R_{AB}^{\mathsf{eff}} = \left( \frac {\kappa} {V_{AB}} \right)^{1/6}
\end{equation}

Here, the value of $\kappa$ is determined from the cross peak intensities for nuclei with known, rigidly constrained distances, such as vicinal protons in aromatic rings.
Experimental effective distances for mobile proton pairs can be computed from the derived value of the $\kappa$ and the measured cross peak intensities of these pairs.
Then, an ensemble-averaged effective distance is computed from an MD trajectory, and force fields can be assessed using a RMSE between the effective distances calculated from the simulations and from the experimental cross peak intensities.
The choice of whether to use $R_{AB}^{\mathsf{slow}}$or $R_{AB}^{\mathsf{fast}}$ to calculate the effective distances from the simulations relies on the same considerations as those described in the previous paragraph.
This approach, too, is less appropriate for intrinsically disordered proteins and  proteins that undergo large conformational changes, as well as for NOESY spectra collected at long mixing times, as these can have non-negligible contributions from spin diffusion.

The third class of approaches uses more detailed models for the dependence of the NOE on intramolecular motions \cite{peter_calculation_2001,vogeli_nuclear_2014} can be applied to compute the dipolar cross relaxation rates and NOESY cross peak intensities directly (Eq. \ref{eqn:redfield} and Appendix \ref{app:spin_relax_redfield}).
Such models may also attempt to account for spin diffusion or for anisotropic molecular tumbling.

In all three approaches, the experimental study may not distinguish between individual hydrogen atoms, either because they are chemically indistinguishable (e.g. the three protons of a methyl group) or because stereospecific hydrogen atoms could not be assigned (e.g. the $\beta$-hydrogens in an amino acid side chain).
These cases may either be omitted from the evaluation or addressed by pseudoatom corrections such as the "center average" approach \cite{wuthrich_pseudo-structures_1983,fletcher_treatment_1996}.

\subsubsection{Spin relaxation}
\label{sub2:spin_relax}

\paragraph{General principles}

A radiofrequency pulse at a nucleus’s resonant frequency tips the spin away from the axis of the external field around which it is precessing, increasing the amplitude of its precession.
It also makes the spins of multiple nuclei precess coherently.
Spin relaxation is the decay of a nucleus’s spin magnetization back to its equilibrium distribution following such a pulse.
This process is typically characterized by two independent exponential decay processes.
Spin-lattice relaxation, or $T_1$ relaxation, is the decay of the component of the nuclear spin magnetization parallel to the external magnetic field, conventionally defined as the $z$ axis.
This decay is caused by the interaction of the aligned spins with their surroundings (often called the lattice despite not being ordered), i.e. an exchange of energy between the spin and the lattice.
This longitudinal component of the magnetization, $M_z(t)$, decays according to

\begin{equation}
\label{eqn:t1_relax}
M_z(t) = M_{z,\mathsf{eq}} - [M_{z,\mathsf{eq}} - M_z(0)] \exp \left( -\frac {t} {T_1} \right)
\end{equation}

\noindent where $M_{z,\mathsf{eq}}$ is the longitudinal component of the magnetization at thermal equilibrium, time $t = 0$ corresponds to the end of the pulse, and $T_1$ is the decay constant for the spin-lattice relaxation.
Spin-spin relaxation, or $T_2$ relaxation, is the decay of the net magnetization transverse to the external magnetic field due to dwindling coherence of the phases of the spins of individual nuclei. The transverse component of the magnetization, $M_{xy}(t)$, decays according to

\begin{equation}
\label{eqn:t2_relax}
M_{xy}(t) = M_{xy}(0) \exp \left( -\frac {t} {T_2} \right)
\end{equation}

\noindent where $T_2$ is the decay constant for the spin-spin relaxation.
For most systems, spin-spin relaxation is faster than spin-lattice relaxation; i.e., $T_1$ > $T_2$.
These decay constants can be measured for NMR-active isotopes in proteins, typically $^{15}N$ in labeled backbone amides or $^{13}$C and $^2$H labeled side chains.
Although protein force fields are usually benchmarked against the spin relaxation times of backbone nuclei, it is also possible to use side chain data \cite{hoffmann_accurate_2018}, though this is generated less often than backbone data.
In addition, heteronuclear NOE (hetNOE) relaxation, which is typically sensitive to fast motions, can be measured.

The Redfield equations \cite{redfield_theory_1965} (see Appendix \ref{app:spin_relax_redfield}) demonstrate that spin relaxation times are related to molecular dynamics via rotational correlation functions of bonds in a complex and magnetic field-dependent manner whose intuitive interpretation is often not straightforward.
To simplify this,  spin relaxation analyses \cite{best_determination_2004,showalter_toward_2007,showalter_validation_2007,maragakis_microsecond_2008,trbovic_structural_2008} typically employ the Lipari-Szabo or model-free approach \cite{halle1981interpretation,lipari_model-free_1982}, where rotations of bonds are assumed to be dominated by two independent types of motion: overall molecular tumbling and intramolecular motions in the molecular frame.
This analysis gives the timescales of protein overall and intramolecular motions, as well as order parameters, $S^2$, characterizing the orientational freedom of bond vectors in the protein frame of reference.
These order parameters range between 0 and 1 \cite{lipari_model-free_1982}, where 0 implies free, isotropic rotation of the bond vector within the molecular frame of reference, and 1 implies a complete absence of such freedom.
Experimental values for the order parameters can be derived for each residue by fitting the value of the order parameter and the time constant for internal motion, along with one additional time constant for molecular tumbling, to the observed spin relaxation rates and NOE enhancements for all residues.
The time constant for molecular tumbling can also be separately fit to the $T_1/T_2$ ratios for all residues in the protein, which are independent of the residue-specific order parameters and time constants for internal motion as long as the internal motions are fast relative to molecular tumbling, an assumption which is typically valid for globular proteins \cite{virtanen_heterogeneous_2020}.

Alternatively, instead of estimating both $T_1$ and $T_2$ values from order parameters using the Lipari-Szabo model, as just discussed, one may estimate $T_2$ from a linear transformation of the effective correlation time for molecular tumbling, obtained from a multi-exponential decay describing the multiple timescales for protein backbone rotational motions \cite{nencini_rapid_2024}.
This can facilitate rapid interpretation of molecular dynamics from spin relaxation times and their comparison with MD simulations.

\paragraph{Evaluation of MD simulations using spin relaxation}

Order parameters and time scales of internal motions computed from MD simulations of folded proteins have been often compared with values extracted from experiments using the Lipari-Szabo model \cite{lipari_protein_1982,best_determination_2004,showalter_toward_2007,showalter_validation_2007,maragakis_microsecond_2008,trbovic_structural_2008,fenwick_classic_2016}.
It is worth noting that this approach cannot be used for proteins with disordered regions because their overall motions cannot be uniquely defined, and their rotational dynamics may not be modeled by a single timescale as assumed in the Lipari-Szabo model.
Nevertheless, spin relaxation times for such systems can be directly calculated from MD simulations by substituting Fourier transformations of multiple rotational correlation functions to the Redfield equations, and then compared with experiments \cite{chen_ab_2018,ollila_rotational_2018,hoffmann_predicting_2020,virtanen_heterogeneous_2020}.
Several codes with slightly different implementation details and varying levels of documentation are available for this: \url{https://github.com/zharmad/SpinRelax} \cite{chen_ab_2018}, \url{https://github.com/ohsOllila/ProteinDynamics} \cite{ollila_rotational_2018}, and \url{https://github.com/nencini/NMR_FF_tools/blob/master/relaxation_times/RelaxationTimes.ipynb} \cite{nencini_probing_2024}.

Sufficiently long simulations \cite{bowman_accurately_2016} can reproduce experimental spin relaxation times to good accuracy without any further corrections \cite{virtanen_heterogeneous_2020}, given that the simulations use a water model which gives an accurate viscosity of water.
In particular, the TIP3P water model \cite{jorgensen_comparison_1983} underestimates the viscosity significantly and therefore leads to unrealistically short time constants for overall protein rotation (tumbling) \cite{wong_evaluating_2008}.
For simulations that use a water model which gives an inaccurate viscosity, overall motion is usually removed mathematically before computing internal relaxation rates \cite{prompers_general_2002,wong_evaluating_2008,anderson_rotational_2012,chen_ab_2018,hoffmann_accurate_2018,ollila_rotational_2018}.

Comparisons of simulated spin relaxation times to experiments can be used to evaluate the accuracy both of protein dynamics and of the conformational ensembles of proteins with long disordered regions \cite{lindorff2012structure,virtanen_heterogeneous_2020}, multi-domain proteins \cite{sandelin_qebss_2024}, and peptides in micelles \cite{nencini_probing_2024}.
Spin relaxation rates are particularly informative for such systems, because they report on dynamics, whereas other observables, such as chemical shifts and scalar couplings, only report time-averaged properties \cite{virtanen_heterogeneous_2020}.
For fully folded proteins, comparisons of spin relaxation times with experiment can be used to evaluate the accuracy of the computed rotational diffusion rates \cite{ollila_rotational_2018}.

The experimental errors are typically between \qty{0.01}{\second} and \qty{0.1}{\second} for $T_1$ and $T_2$ values and, when using the full Redfield equations (Appendix \ref{app:spin_relax_redfield}), between 0.05 and 0.1 for the backbone amide heteronuclear NOE values \cite{hoch_biological_2023}.
These experimental uncertainties are smaller than the usual statistical uncertainty in MD simulations \cite{sandelin_qebss_2024}.

\subsubsection{Paramagnetic relaxation enhancement}
\label{sub2:pre}

\paragraph{General principles}

Paramagnetic relaxation enhancement (PRE) is an increase in the NMR relaxation rate of a nuclear spin due to its dipolar interactions with unpaired electrons at a paramagnetic site; i.e., at an atom with an unpaired electron \cite{clore_elucidating_2007,clore2009theory}.
Although the dipolar cross relaxation that causes PRE is essentially the same as that which causes NOE, PRE ranges over much longer distances, to as far as \qty{2.5}{\nano\meter}, because the magnetic moment of an unpaired electron is much larger than that of a nucleus.
Some metal atoms in metalloproteins are paramagnetic and generate measurable PREs.
For other proteins, paramagnetic sites can be artificially introduced by using chemical reactions to attach extrinsic labels, called spin-labels.
This is usually done by engineering proteins to have just a single, reactive cysteine residue which can then be reacted with a nitroxide-containing compound, such as MTSL, or a chelating agent, such as an EDTA derivative, carrying a paramagnetic metal.
Although it is possible to measure the PRE for longitudinal NMR relaxation rates ($R_1=T_1^{-1}$), most applications focus on transverse relaxation rates ($R_2=T_2^{-1}$), and these will be the focus of this section.
PRE depends on both the distribution of distances and the timescales of motion of the protein, along with the location and dynamics of the spin-label.
The strength, range, and strong dependence on distance make PRE particularly suitable to detect and quantify transient and low-probability interactions.

In a typical PRE experiment, one measures the transverse relaxation rates of various groups in both the spin-labeled (paramagnetic) protein and in the same protein without the spin label (the diamagnetic protein).
The PRE for nuclei across the protein then is obtained from the difference between the spin relaxations of these two measurements.
Alternatively, because the PRE leads to line-broadening, one may estimate the PRE from the ratio of the intensities (peak heights) in e.g. heteronuclear single quantum correlation spectra of the paramagnetic and diamagnetic samples.
However, although this has commonly been done and may lead to useful insights, the analysis of such data comes with additional uncertainty and assumptions so that one should, if possible, measure the PRE via relaxation rates \cite{clore2009theory}.
When a nitroxide spin-label is used, the diamagnetic protein can be generated simply by reducing the nitroxide with ascorbic acid.
It is possible to measure PREs of various nuclei and chemical groups, but they are most commonly measured at backbone amides.

\paragraph{Evaluation of MD simulations using PRE}

Like NOEs (Section \ref{sub2:noesy}) and spin relaxation rates (Section \ref{sub2:spin_relax}), PREs depend on the intramolecular motions of proteins in a complex manner (see Appendix \ref{app:pre_solomon}), and approaches used to handle intramolecular motions for NOEs and spin relaxation rates are applicable to PREs as well.
Specifically, the PRE can be interpreted as providing a probe of the ensemble-averaged distances between probed spins and the spin label.
An effective ensemble-averaged effective distance calculated from an MD trajectory, such as $R_{AB}^{\mathsf{slow}}$ (Eq. \ref{eqn:noe_R_slow}), can be compared to distances calculated from experimental PREs.
Alternatively, a model for the intramolecular motion of the protein, such as the Lipari-Szabo approach, can be used to compare order parameters or PREs directly to the corresponding experimental values.

The PRE has been used most extensively to benchmark MD simulations of disordered proteins with weak and/or transient interactions \cite{piana_water_2015,robustelli_developing_2018}, but it can also be used to probe transient interactions between folded proteins and between proteins and nucleic acids, and could thus be used to benchmark simulations of such systems.
PREs have also been used to parameterize coarse-grained force fields for disordered proteins \cite{norgaard2008experimental,tesei2021accurate}.
Often, PRE experiments involve measurements on proteins labeled at multiple sites, one at a time, to get a global view of the structural dynamics of a protein.
Each experiment probes the distance between the spin-label site and all backbone amide protons.
Thus, one is faced with the challenge of assessing a simulation of a protein based on all these measurements.

When calculating PREs from MD simulations and comparing them to spin-label experiments, one must decide how to model the spin-labels.
One approach is to simulate each variant of the protein with its covalently linked spin label.
This requires generating force field parameters for the covalently modified protein and often also requires mutating the sequence to exclude/include cysteine residues to match the experimental protein constructs.
One must then repeat the simulation for each spin-label site of interest.
An easier alternative is to simulate the unlabeled wild-type protein and calculate distances between protein atoms as proxies for the distances between the spin-label and amide protons.
For example, one may use an atom in the sidechain of the reference (wild-type) sequence as a proxy for the location of the unpaired electron.
However, this approach risks missing genuine effects of the spin-labels on the protein’s dynamics.
A compromise between these two extremes is to perform simulations of the wild-type protein and then model the spin-label onto this simulation using a rotamer library developed to describe the structural preferences of the label.
Such rotamer libraries are available for the commonly used MTSL spin-label \cite{polyhach_rotamer_2011}.
For applications to large MD simulations, placing the spin-labels and sampling the rotamers may be achieved by tools such as Rotamer-ConvolveMD in the MDAnalysis package \cite{michaud-agrawal_mdanalysis_2011,gowers_mdanalysis_2016} and DEER-PREdict \cite{tesei_deer-predict_2021}, with the latter also implementing calculations of the PREs from the simulations.
Which of these approaches to choose depends in part on the desired accuracy and whether there is experimental evidence that the spin-label itself introduces a change in the conformational ensemble.

While PREs can be measured relatively accurately, a number of issues complicates their use in assessing force field accuracies.
First, unless the spin-label is modeled explicitly, there may be uncertainty regarding the degree to which the spin label affects the protein \cite{sasmal2017effect,tesei_deer-predict_2021}.
Second, PREs are often probed indirectly via measurements of intensity ratios, limiting accuracy and interoperability \cite{clore2009theory}.
Finally, unless the timescales of motions are calculated directly from the simulations \cite{xue2011motion}, calculations of PREs require either estimating \cite{tesei_deer-predict_2021} or fitting \cite{tesei2021accurate} the time-scales, approaches that introduce uncertainty.
Due to these complications, the root mean square error between simulated and experimental values (PREs directly or, indirectly, effective distances or order parameters) should be compared between simulations of the same protein with different force fields but should not be compared between simulations of different proteins.

\subsection{NMR datasets}
\label{sub:nmr_data}

Many NMR datasets for proteins and peptides are available in the literature and in the Biological Magnetic Resonance Bank (BMRB) \cite{ulrich_biomagresbank_2008,romero_biomagresbank_2020}.
We focus here on datasets, listed in Table~\ref{tab:datasets}, that appear particularly useful for benchmarking force field parameters because they largely meet the following criteria: 1) the data are available in a machine readable format; 2) estimates of experimental uncertainty are included; 3) a diversity of structural motifs are present, including different secondary and tertiary structure elements and disordered regions; 4) the proteins are small enough that relatively short simulations suffice to provide well-converged estimates of the NMR observables.
Note that many NMR observables for proteins were collected under low pH conditions, so it is essential to assign protonation states that are consistent with the environment of the experimental measurements (see Section \ref{sub:best_practices_setup}).
This can complicate force field benchmarks, because some force fields may not have parameters for protomers of the standard amino acids, especially for a protonated C terminus.

\subsubsection{Beauchamp short peptides and ubiquitin}
\label{sub2:beauchamp}

Beauchamp et al. \cite{beauchamp_are_2012} curated from the literature and Biological Magnetic Resonance Bank 524 NMR chemical shifts and scalar J-couplings for 19 capped 1-mers of the form Ace-X-Nme, where X are non-proline amino acids; 11 uncapped 3-mers; tetraalanine; and the protein ubiquitin.
The short peptides in this dataset provide an opportunity to assess the bacbone preferences of amino acid residues in the absence of a defined secondary structure.
It is important for a force field to capture these preferences in order to accurately predict the conformational distributions of flexible loops in folded proteins and of unfolded and intrinsically disordered proteins.

\subsubsection{Designed $\beta$-hairpins and Trp-cage miniproteins}
\label{sub2:designed_beta}

Many groups have used NMR to characterize the solution structure, stability, and dynamics of designed beta-hairpin sequences \cite{blanco_short_1994,ramirez-alvarado_novo_1996,de_alba_turn_1997,maynard_origin_1998,stanger_rules_1998,cochran_tryptophan_2001,ramirez-alvarado_elongation_2001,pastor_combinatorial_2002,russell_stability_2003,skelton_hairpin_2003,fesinmeyer_enhanced_2004,espinosa_autonomously_2005,andersen_minimization_2006,honda_crystal_2008,kier_stabilizing_2010,scian_mutational_2013,jimenez_design_2014,anderson_nascent_2016,morales_design_2019,richaud_folding_2021,peintner_pushing_2022} and miniproteins \cite{struthers_design_1996,mcknight_nmr_1997,struthers_design_1998,imperiali_uniquely_1999,vermeulen_solution_2004,cornilescu_solution_2007,gronwald_evolutionary_2008} in a variety of solution conditions.
One series of studies of Trp-cage miniprotein sequences \cite{neidigh_designing_2002,lin_helical_2004,barua_trp-cage_2008,scian_crystal_2012,byrne_circular_2013,graham_reversing_2019} includes NOE restraints for four Trp-cages (PDB entries 1L2Y \cite{neidigh_designing_2002}, 2JOF \cite{barua_trp-cage_2008}, 2M7D \cite{byrne_circular_2013}, 6D37 \cite{graham_reversing_2019}), folding rates measured by NMR resonance line-broadening due to folded/unfolded-state exchange \cite{scian_mutational_2013,byrne_folding_2014}, and temperature-dependent chemical shift deviations (CSDs) for dozens of related sequences measured in the same solution conditions.
These measurements offer a high-quality benchmark set that reports on how mutations perturb folding.
Molecular simulations are now routinely able to access the microsecond timescales required to make accurate comparisons with these observables.

\subsubsection{Stroet folded proteins}
\label{sub2:stroet}

Stroet et al curated a dataset of 52 high resolution protein structures, ranging in size from 17 to 326 residues \cite{stroet_validation_2024}.
All of the proteins are monomeric in solution, and none contain ligands or cofactors.
13 protein models, ranging in size from 17 to 129 residues, were derived from NMR experiments.
For nine of these, NMR observables including upper bounds for interatomic distances derived from NOE data, $^3J$-couplings, and RDCs are available, either from the PDB or from the Biological Magnetic Resonance Data Bank (BMRB) \cite{ulrich_biomagresbank_2008,romero_biomagresbank_2020,hoch_biological_2023}. This dataset, which is available from the Australasian Computational and Simulation Commons (ACSC) Molecular Simulation Data Repository \cite{mark_australasian_2022} at \url{https://molecular-dynamics.atb.uq.edu.au/collection/protein-force-field-validation-set}, was used to validate backbone torsional parameters \cite{diem_hamiltonian_2020} and to study the impact of nonbonded cutoff schemes on experimental observables \cite{diem_effect_2020}.

In preparing the dataset, Stroet et al curated the NOESY studies that provide interatomic distances derived from NOESY intensities but not the intensities themselves.
Any pseudoatom corrections included in the experimental data were first removed, and then the pseudoatom corrections proposed by W\"{u}thrich \cite{wuthrich_nmr_1986} were applied.
Pseudoatom corrections were applied in this way to six of the nine proteins with NOE distance restraints.
For two of the remaining three proteins, no information on any pseudoatom corrections to the NOE-derived upper bounds could be found, and so the NOE restraints reported for these proteins may be less rigorous.
For the third, mercury binding protein from \textit{Shigella flexneri} (PDB-code 1AFI), a large number of violations in NOEs assigned to Phe47 were observed in initial simulations.
Inspection of the structure and the violations points to wrongly assigned explicit phenyl ring hydrogens HE1/2 and HD1/2.
Replacing the assignment of these protons on the symmetrical phenyl ring by pseudoatoms HE* and HD*, and adding the appropriate pseudoatom corrections \cite{wuthrich_nmr_1986} removed all violations above 0.05 nm \cite{stroet_validation_2024}.

\subsubsection{Mao folded proteins}
\label{sub2:mao}

This is a collection of 41 folded proteins for which both X-ray structures and NMR data, comprising backbone chemical shifts and NOESY intensities, have been measured by the Northeast Structural Genomic Consortium \cite{mao_protein_2014}.
These data have been used to assess the accuracy of NMR structures, X-ray structures, and Rosetta refinements \cite{mao_protein_2014}, and to compare the accuracy of MD simulations run with various force fields \cite{robustelli_developing_2018,piana_development_2020}.

\subsubsection{Robustelli folded and disordered proteins}
\label{sub2:robustelli}

Robustelli, Lindorff-Larsen, and co-workers assembled a dataset (\url{https://github.com/paulrobustelli/Force-Fields}) that spans 21 proteins and peptides with over 9,000 experimental data points and probes the ability of a protein force field to simultaneously describe the properties of folded proteins, weakly structured peptides, and disordered proteins with a range of residual secondary structure propensities \cite{lindorff-larsen_improved_2010,lindorff2012systematic,piana_water_2015,robustelli_developing_2018,piana_development_2020}.

The dataset contains four fully folded proteins (ubiquitin, GB3, hen egg white lysozyme (HEWL), and bovine pancreatic trypsin inhibitor) with extensive NMR data, including backbone and sidechain $J$-couplings, RDCs, and backbone and side chain spin relaxation order parameters.
It also includes calmodulin and the bZip domain of the GCN4 transcription factor, both of which contain folded and flexible components.
Calmodulin, which comprises two folded domains connected by a flexible linker, probes the ability of a force field to simultaneously describe the flexibility of the linker region, the stability of folded domains, and the propensity of the folded domains to associate.
The NMR data for calmodulin comprise chemical shifts and RDCs.
The bZip domain of the GCN4 transcription factor is a partially disordered dimer with an ordered, helical, coiled-coil dimerization domain, for which NMR chemical shifts and backbone amide spin-relaxation parameters are available. 

The dataset also includes nine proteins that are disordered under physiological conditions and for which extensive sets of NMR data are available.
These test the ability of a force field to accurately describe the dimensions and secondary structure propensities of intrinsically disordered proteins.
The proteins range in size from 40 to 140 amino acids, which was important, as a number of force fields that produced reasonable dimensions for proteins containing <70 amino acids produced conformations that were substantially over-collapsed for longer sequences.
The available NMR data for these disordered proteins include chemical shifts, RDCs, backbone $J$-couplings, and PREs.
Scalar couplings of the disordered Ala5 peptide were also included. 
Code to calculate NMR and SAXS data for five IDPs in this benchmark (A$\beta$40, drkN SH3, ACTR, PaaA2 and $\alpha$-synuclein) is included in recent manuscript describing a maximum entropy reweighting approach to calculate conformational ensembles of IDPs from experimental data\cite{borthakur2024determining} (\url{https://github.com/paulrobustelli/Borthakur_MaxEnt_IDPs_2024}).  

The Robustelli dataset is enriched by a number of non-NMR data, including radii of gyration obtained by various experimental methods and data on the temperature-dependent stability of fast-folding proteins
The dataset has also been expanded \cite{piana_development_2020} to include the free energies of association of 14 protein-protein complexes, the osmotic coefficients of 18 organic and inorganic salts, the position of the first peak of the radial distribution functions of seven ion-water and ion-ion pairs, comparisons of Ramachandran distributions of blocked amino-acids in water, Ramachandran distributions obtained from X-ray coil libraries, Lifson-Roig helix extension parameters for the 20 amino acids (estimated from NMR studies of Ace-(AAXAA)3-Nme peptides), the folding free energies of mutants of 22 mutants of Trp-cage, the folding enthalpies of 10 fast-folding proteins, the Kirkwood-Buff integrals of ethanol water-mixtures, and the melting curves of the Trpzip1 and GB1 $\beta$-hairpin-forming peptides.

\subsubsection{Spin relaxation datasets}
\label{sub2:spin_relax_datasets}

Spin relaxation data have been reported for a large number of proteins; see for example a comprehensive review of spin relaxation measurements published before 2006 \cite{jarymowycz_fast_2006}.
Furthermore, the BMRB contains spin relaxation data for approximately 350 proteins.
\cite{ulrich_biomagresbank_2008,romero_biomagresbank_2020,hoch_biological_2023}.
Although most have not so far been analyzed in conjunction with molecular simulations, there are some useful examples of such analyses.

First, the membrane-bound, bacterial TonB proteins possess a long, disordered region that links their C-terminal domain to their transmembrane N-terminus.
Spin relaxation times have been measured for the C-terminal domain with varying linker lengths (HpTonB194-285, HpTonB179-285, HpTonB36-285) \cite{ciragan_nmr_2020}, and these data can be used to evaluate the conformational ensembles predicted by MD simulations.
In one such study \cite{virtanen_heterogeneous_2020}, simulations with the Amber ff03ws force field reproduced the experimental spin relaxation times of partially disordered fragments of TonB as well as Engrailed 2 (see below), while CHARMM36m and Amber ff99-ILDN gave less accurate results, apparently because they yielded overly collapsed conformational ensembles.
Interestingly, however, the three force fields gave similar accuracy for chemical shifts, presumably because of degeneracy in chemical shifts with respect to the structural features important for describing this ensemble.
Second, spin relaxation data at multiple magnetic fields are available for the partially disordered 143–259 region of the Engrailed 2 transcription factor \cite{khan_distribution_2015}.
This region is highly conserved and is involved in the binding of transcriptional regulators.
Simulations with the Amber ff03ws force field yielded good agreement with the experimental data, except for serine and aspartate residues \cite{virtanen_heterogeneous_2020}.

Spin relaxation times have also been used to find the best ensembles among diverse simulation data for four different multi-domain proteins using the quality evaluation based simulation selection (QEBSS) approach \cite{sandelin_qebss_2024}, and to characterize dynamics of six different peptide-micelle systems in conjunction with MD simulations \cite{nencini_probing_2024}.

\subsubsection{NMR-derived salt bridge stabilities}
\label{sub2:salt_bridge}

In some cases, NMR data can be used to measure secondary properties which can in turn be compared with simulations.
For example, the thermodynamic stability of weak, solvent-exposed salt bridges can be assessed by monitoring NMR chemical shifts as a function of pH.
One such study examined three potential salt bridges in the context of a folded protein, the B1 domain of protein G (GB1) \cite{ahmed_how_2018}.
These involve lysine-carboxylate ionic interactions and were identified from crystal structures.
The stabilities of these potential salt bridges were assessed via $^{15}$N and $^1$H chemical shifts and the hydrogen-deuterium exchange rates of the lysine ammonium group during titration of the carboxylates \cite{tomlinson2009characterization}.
The NMR data indicate that two of the salt-bridges are not formed in solution while the third is only weakly formed.
Interestingly, most force fields tested overestimated the stability of the salt bridges \cite{ahmed_how_2018}, a result also reported by previous studies \cite{piana_how_2011,debiec_evaluating_2014} looking at the association constants of oppositely charged amino acids in water, where the experimental data were obtained by potentiometric titration rather than NMR.
Here, the results were significantly improved by atomic charge derivation strategies that implicitly incorporate solvent polarization \cite{debiec_further_2016} and by the use of the more expensive, polarizable CHARMM Drude-2013 \cite{lopes_polarizable_2013} and AMOEBA force fields \cite{shi_polarizable_2013}.

\section{Room-temperature protein crystallography}
\label{sec:xtal}

The earliest protein crystal structures were determined by X-ray diffraction for specimens at or near room temperature \cite{muirhead_structure_1963}.
Later, methods of working with protein crystals at low temperature were widely adopted for their practical advantages, such as slowed radiation damage of the crystal specimen \cite{fischer_macromolecular_2021,thorne_determining_2023}.
More recently, however, there has been a revival of interest in room temperature protein crystallography, because it gives insight into molecular motions that are quenched at cryogenic temperatures \cite{fraser_hidden_2009,keedy_crystal_2014,fischer_macromolecular_2021,thorne_determining_2023}.
At cryogenic temperatures, there is little motion to simulate, and it is also not clear that the conformational variability observed at cryogenic temperatures corresponds to a well-defined thermodynamic ensemble at any particular temperature \cite{keedy_crystal_2014,bradford_temperature_2021}.
In contrast, the conformational variability observed by RT crystallography is more clearly attributable to room temperature thermal motions.
Thus, although most crystallographic data available in the Protein Data Bank (PDB) were measured at cryogenic temperatures, the PDB contains a growing number of room temperature structures.
Neutron diffraction crystallography (Section \ref{sub2:neutron}), which also can be carried out at room temperature, goes beyond X-ray crystallography because it can resolve the positions of hydrogen atoms in protein structures.
However, there are not many neutron diffraction structures in the PDB as there are not many suitable neutron sources where these studies can be done, and the method requires larger crystals.

\begin{figure}[t]
    \centering
    \includegraphics[width=\linewidth]{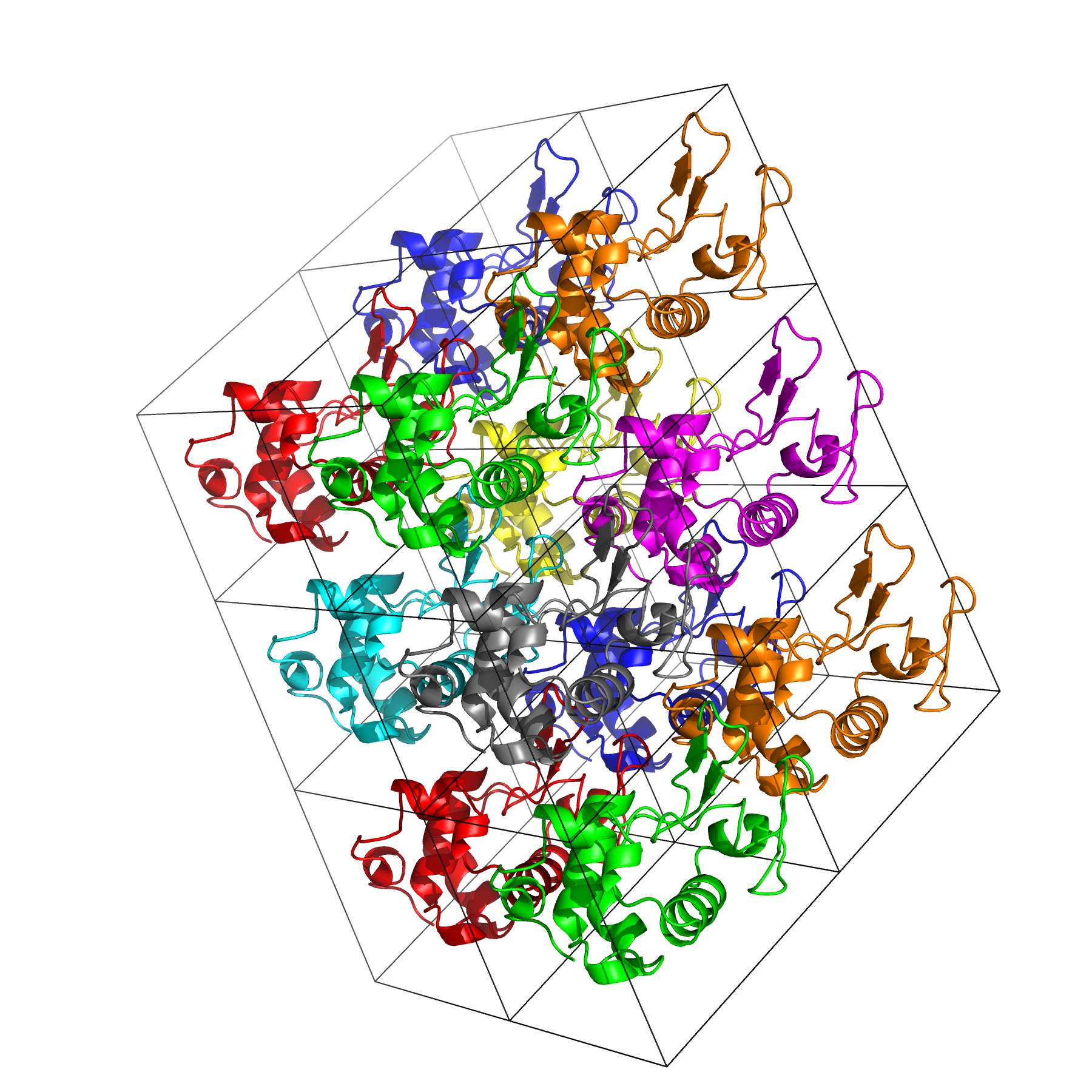}
    \caption{A 3 $\times$ 2 $\times$ 2 supercell containing 12 independent copies of hen egg white lysozyme (4LZT \cite{walsh_refinement_1998}), each colored differently from its neighbors.
    Black lines represent cell boundaries of the triclinic unit cell.}
    \label{fig:supercell}
\end{figure}

Water typically occupies about 50\% of a protein crystal’s volume \cite{altan_learning_2018}, so crystallized proteins are usually quite well-solvated.
Nonetheless, the conformational distribution of a crystallized protein is likely to differ from that of the same protein in solution, due to protein-protein contacts, perturbations of water structure, and the presence of cosolutes added to facilitate crystallization.
Therefore, when one uses crystallographic data to benchmark molecular simulations, one should simulate the crystal, rather than the protein in solution.
Crystal simulations with periodic boundary conditions can in principle be used to simulate a single unit cell, but this does not allow variations between adjacent lattice positions.
Therefore, it is preferable to simulate a supercell comprising multiple unit cells which can sample such lattice variations (Fig. \ref{fig:supercell}).
This requirement adds complexity and can lead to larger and hence slower simulations than those used with NMR benchmark data, depending on whether the size of the supercell exceeds the size of the simulation box for the solvated system.
At the same time, the extra copies of the protein in supercell simulations often can be included in computing statistical averages, enabling some analyses to be performed using shorter simulation durations than for conventional solution simulations. 
Different proteins, and even different crystal structures of the same protein, have different levels and characters of both conformational variation and experimental error.
Therefore, much as for NMR benchmarking, force fields should be compared against a single crystallographic dataset, rather than attempting to compare force fields based on benchmarks against different crystal structures.
Methods of simulating protein crystals have been discussed in recent reviews \cite{cerutti_molecular_2019,wych_molecular-dynamics_2023}.

This section reviews crystallographic observables that are relatively well developed for use in benchmarking protein simulations, touches on the additional topic of diffuse scattering, and lists crystallographic datasets that are well-suited for benchmarking.

\subsection{Crystallographic observables}
\label{sub:xtal_obs}

\subsubsection{Bragg diffraction data}
\label{sub2:bragg}

\paragraph{General principles}

X-ray scattering from an ideal crystal would be focused into sharp spots on the detector, located at positions that are generally called the Bragg positions, and the intensities at these positions are sensitive to the time- and space-average of the electron density.
Diffuse scattering, which reports on correlated motions and crystalline disorder \cite{meisburger_x-ray_2017}, is discussed in Section \ref{sub2:diffuse}.
The Bragg intensities are related to the mean unit-cell electron density through the complex structure factors, i.e. the Fourier transform of the unit-cell electron density distribution.
In particular, the Bragg intensities are proportional to the squared amplitudes of the structure factors.
The phases are also needed to compute the electron density, and in the absence of an anomalous dispersion signal from which experimental phases can be obtained, the phases must be obtained from an atomic model.
The electron density therefore is not typically an "experimental observable" in the usual sense.
Still, for a well-refined model, with good resolution and an R-factor below 15\% or so, it is common to refer to the "observed electron density".

\paragraph{Evaluation of MD simulations using Bragg diffraction data}

The simplest way to assess the accuracy of a simulation of a protein crystal is to compare the average structure from the simulation to the crystallographic model structure available in the Protein Data Bank (PDB) \cite{bernstein_protein_1977,berman_protein_2000,rose_rcsb_2015}.
This intuitive and quick comparison may be useful when there are large differences between two force fields of interest.
In addition, the focus on a single structure may help to identify specific problems in the force field, such as insufficient stability of alpha-helices or side-chains preferring uncommon rotamers, and thus can yield feedback to improve it.
To assess protein backbone agreement, RMSD of C$\alpha$ atoms or of all backbone heavy atoms can be used.
To assess protein side-chain agreement, RMSD of side-chain atoms, dihedral-angle agreement within an angular threshold, or match between defined rotamers \cite{dunbrack_backbone-dependent_1993,lovell_penultimate_2000}, can be used.
Note that a rigorous measure of RMSD must account for symmetries; for example, a \ang{180} rotation of the $\chi_2$ angle of a phenylalanine in the model should not alter the computed RMSD.
The possibility of alternate conformations (Section \ref{sub2:alt_confs}) should also be factored in.
While this general approach is widely used and simple to implement, the structural models used as the basis for the comparison are typically derived by refinement informed by an empirical force field.
Therefore, comparisons against the structural model risk assessing, at least in part, the agreement of the force field of interest with the force field used during refinement, rather than agreement with experimental data.

A second approach is to compute the mean electron density from the simulation \cite{grosse-kunstleve_computational_2002,wall_methods_2009,winn_overview_2011,case_amber_2022} and compare it with the crystallographic electron density.
This avoids the intermediary of a structural model inferred from the crystallographic density.
Note that the observed electron density is averaged over time and over many unit cells of the crystal, and hence reflects disorder in the experimental system.
Because the density is often sharply peaked around atomic positions, standard methods of comparing density, such as the Pearson correlation, can be more sensitive to small, local differences than is desired for assessing the relative accuracy between force fields, although these issues may be addressed by approaches such as local model-based map alignments \cite{pearce_multi-crystal_2017}.
Existing crystallographic software, such as EDSTATS in CCP4 \cite{winn_overview_2011} and \verb|phenix.real_space_correlation| in PHENIX \cite{liebschner_macromolecular_2019}, can be used to compare electron density maps.

Other methods of comparing simulations to crystallographic data are also possible.
For example, one could compute the mean electron density from the simulation using mdv2map in AmberTools with CCP4 \cite{winn_overview_2011,case_amber_2022} or xtraj in LUNUS with CCBTX \cite{grosse-kunstleve_computational_2002,wall_methods_2009}, use crystallographic software to refine an atomic model into this density, and then use structural metrics such as atomic RMSD to compare this model with the experimental structure model \cite{wych_molecular-dynamics_2023}.
Alternatively, one could directly compare computed and measured Bragg intensities.
This properly takes care of disorder, but can be more difficult to interpret, as modest deviations (by an RMSD less than \qty{0.05}{\nano\meter}) of the protein away from the "correct" structure can greatly reduce the real-space correlations between computed and experimental diffraction intensities \cite{wall_internal_2018}.
In such cases, other measures of accuracy than correlation may be more useful.
One may also focus on the accuracy of protein-protein contacts at interfaces, as these depend on non-covalent and solvent-mediated interactions that are important to get correct.
Most crystal contact analysis methods, such as the PISA software \cite{krissinel_inference_2007}, use a single average structure as input, so it is difficult to account directly for disorder.
As with the comparison of average protein structures discussed above, this sort of comparison may be most useful when there are relatively large differences between the results from different proposed force fields.

\subsubsection{Debye-Waller factors (B-factors)}
\label{sub2:b_factors}

\paragraph{General principles}

The Debye-Waller factor of an atom, also known as its B-factor or temperature factor, is closely related to the atom’s fluctuations about its modeled position.
B-factors are, in principle at least, related to the mean square displacement of the atom, $\langle u^2 \rangle$, by the equation

\begin{equation}
\label{eqn:b_factor}
B = \frac {8} {3} \pi^2 \langle u^2 \rangle
\end{equation}

\noindent However, B-factors can also have contributions from various sources of error, and are sometimes viewed as "slop factors" that absorb other errors in the model to improve the fit to the diffraction data.
They may result from incorrect assignments of the identities of atoms of similar atomic number, i.e. K vs. Ca, and may also reflect, in a nonspecific manner, incorrect modeling of side chains, alternate conformations, and disordered regions.
Debates continue between crystallographers about whether a region that exhibits signs of disorder should be left unmodeled, modeled with zero occupancy, or modeled with B-factors allowed to refine to high values; as a result, their meaning can vary between different crystal structures.
However, when derived from well-validated structural models \cite{williams_molprobity_2018} that are modeled correctly and carefully interpreted, they provide a meaningful measure of the mobility of atoms within a protein structure.
For structures with resolutions better than \qty{0.15}{\nano\meter}, it is often appropriate to refine anisotropic B-factors, such that each atom is assigned a tensor of six parameters that define a three-dimensional Gaussian distribution of atomic fluctuations.

\paragraph{Evaluation of MD simulations using Debye-Waller factors}

Given a simulation of a protein crystal, it is straightforward to compute the mean square displacement of each atom from its mean position and thus obtain computed B-factors; a modest amount of additional complexity is involved in computing anisotropic B-factors.
Alternatively, one may, again, refine a structural model into the electron density computed from the simulation, and compare the resulting simulated B-factors with the experimental B-factors.
Depending on the software used for refinement, constant offsets may be applied to reported B-factors using methods that do not distinguish contributions from molecular motion from other contributions \cite{liebschner_macromolecular_2019}.
Constant offsets therefore are often ignored in B-factor profile comparisons by using, e.g. a Pearson correlation to assess the agreement. Nevertheless, it is sometimes possible to account even for the constant offset using MD simulations \cite{wall_internal_2018}.

\subsubsection{Alternate conformations}
\label{sub2:alt_confs}

\paragraph{General principles}

In contrast to B-factors, which in principle describe fluctuations within a local energy well, alternate conformations explicitly describe jumps between discrete conformational states in separate local energy wells.
For example, alternate conformations---also known as alternative conformations, alt confs, alternate locations, or altlocs---can be used to model amino-acid side chains that switch between different rotamer \cite{dunbrack_backbone-dependent_1993,lovell_penultimate_2000} conformations.
Alternate conformations are a promising avenue for comparison to and improvement of simulations.
In a protein structure file, an alternate conformation is given as an additional set of coordinates for a group of atoms (often a residue or series of residues) and marked with a single-character identifier (A, B, C, etc.) that is unique within the file.
Each alternate conformation is also assigned a partial occupancy (i.e. probability) from 0 to 1 that is determined by the crystallographic refinement.
For covalently bound atoms, like those of proteins, the occupancies of all alternate conformations for a given atom typically sum to one.
For molecules that are not covalently bound to the protein, such as ligands and water molecules, the occupancies may sum to less than one.
A crystal structure model that contains alternate conformations is termed a multiconformer model.

Alternate conformations are often left unmodeled in crystal structure models even when they are evident in the electron density maps \cite{lang_automated_2010}.
Such missing alternate conformations can be modeled in an automated and unbiased manner with tools such as the qFit software \cite{riley_qfit_2021}, and MD simulations of protein crystals can be used to find alternate conformations that are missed by such automated methods \cite{wych_molecular-dynamics_2023}.
Alternate conformations are significantly more prevalent at RT than at cryogenic temperatures (cryo) \cite{fraser_accessing_2011}, and some alternate conformations that are observed only at RT and not at cryo are critical to biological function \cite{fraser_hidden_2009} (see Section \ref{sub2:cypa}) and can modulate ligand binding in important ways \cite{bradford_temperature_2021}.
Thus, RT crystallography can reveal biologically relevant conformational heterogeneity.

When different parts of a biomolecule have alternate conformations, these conformations may not be mutually independent.
For example, a rearrangement of one part of a protein might restrict which conformations its neighbors can adopt.
However, the RCSB PDB format for crystal structure models does not provide a mechanism for specifying which alternate conformations are physically compatible with one another, other than the A, B, C, etc. identifiers themselves.
This bookkeeping issue creates ambiguity in situations where, for example, residue X with conformations A and B is near residue Y with conformations A, B, and C.
In such a case, it is unclear which conformations of residue X are energetically compatible with conformation C of residue Y.
The corresponding alternate conformations might be coupled, uncoupled, or partially coupled.

\paragraph{Evaluation of MD simulations using alternate conformations}

The fractional occupancies of alternate conformations inferred from RT crystallography can be readily compared with the corresponding results from a simulation.
Conformational probabilities can be extracted from simulations based on the occupancy of rotameric energy wells, e.g. the rotamer name strings \cite{lovell_penultimate_2000} that are output by software such as the model validation suite MolProbity \cite{williams_molprobity_2018} or the related tool \verb|phenix.rotalyze| from PHENIX \cite{liebschner_macromolecular_2019}.
A figure of merit for the ability to reproduce alternate conformations could be the root-mean-square error between rotamer probabilities from simulations and experimental occupancies of alternate conformers, although future studies could explore more sophisticated approaches.
Note that there are datasets where occupancies have been shown to vary with temperature \cite{bradford_temperature_2021}.
In addition, the coordinates and occupancies of alternate conformations can be combined with B-factors to calculate so-called crystallographic order parameters, which may be compared with experimental NMR order parameters \cite{fenwick_integrated_2014}; a similar framework could be useful for comparing RT crystallographic models to simulations.

When comparing computed and observed populations, it is important to bear in mind that small changes in the free energy difference between two conformations will lead to large changes in occupancy if the free energy difference is near zero, but essentially no change if the free difference is far from zero.
This is because, given two conformers A and B, the probability of being in state A is
\begin{equation}
\label{eqn:alt_confs}
p_A = \left( 1 + \exp \left(- \frac {\Delta G_{AB}} {R T} \right) \right )^{-1}
\end{equation}
\noindent where $\Delta G_{AB}$ is the free energy change on going from A to B; and $p_B = 1-p_A$, assuming only two accessible conformations.

\subsubsection{Diffuse scattering (non-Bragg reflections)}
\label{sub2:diffuse}

Diffuse or continuous scattering refers to the cloudy, streaked, speckled, halo-shaped, or otherwise patterned weak scattering that lies between the Bragg peaks.
In contrast with the Bragg scattering, which is associated with correlations in the mean electron density, the diffuse scattering is associated with spatial correlations in the deviation of the density from the mean.
Thus, much as the Bragg peaks can be used to model the mean structure of the molecules within the unit cell, the diffuse scattering can, at least in principle, be used to model the coupling of variations in atom positions.
There have been several protein crystal MD simulations of diffuse scattering, and it is straightforward to compute diffuse intensities from a collection of simulation snapshots \cite{faure_correlated_1994,clarage_sampling_1995,hery_x-ray_1998,meinhold_fluctuations_2005,wall_conformational_2014,wall_internal_2018,wych_liquid-like_2019,meisburger_diffuse_2020,meisburger_robust_2023}.
Although a limited study did show that the force field can influence the simulated diffuse scattering \cite{wych_liquid-like_2019}, it is not yet clear what sort of variation one should expect.
Future crystal simulations should help clarify whether diffuse scattering data can be useful for benchmarking force fields.

\subsubsection{Hydrogen coordinates from neutron diffraction}
\label{sub2:neutron}

Protein X-ray crystallography generally resolves the coordinates only of atoms with atomic number greater than one; hydrogens are visible only in well-ordered regions of ultra-high resolution structures.
This is because X-rays are scattered by electrons, and a hydrogen atom, with only one electron, scatters only weakly.
In contrast, neutrons are scattered by atomic nuclei, and hydrogen atoms have high scattering cross-sections.
As a consequence, neutron diffraction protein crystallography can provide experimental information on hydrogen atom positions and protonation states \cite{blakeley_neutron_2009,ashkar_neutron_2018}.
These data are valuable because hydrogens make up nearly half the atoms in proteins and can play critical functional and structural roles \cite{word_asparagine_1999,word_visualizing_1999}.
Thus, neutron diffraction studies are uniquely suited to discern hydrogen bonding patterns \cite{chen_direct_2012,nakamura_newtons_2015}, the orientations of solvents and side chain groups \cite{blakeley_15-k_2004,wall_biomolecular_2019}, and the protonation states of critical catalytic amino acids---such as the histidine in the catalytic triad of a serine protease \cite{kossiakoff_neutron_1980,kossiakoff_direct_1981}.
By the same token, they provide distinctive information to benchmark protein force fields.

Neutron diffraction is generally weak, and therefore requires large crystals and long data collection times.
In addition, there are only a few suitable neutron sources in the world.
However, macromolecular neutron crystallography has become more practical in recent years with the commissioning of new, higher flux neutron sources, as well as improved techniques for crystal growth and molecular biology techniques for producing large amounts of purified proteins.

\subsection{Protein crystallography datasets}
\label{sub:xtal_data}

Here, we highlight protein crystallography datasets that are well-suited to benchmark force field accuracy, prioritizing the following features: 1) resolution better than \qty{0.12}{\nano\meter}, so that observable data have low uncertainty; 2) relatively unambiguous assignment of protonation states; 3) absence of ligands or co-factors which might conflate the accuracy of the small molecule force field with that of the protein force field; 4) diversity of secondary and tertiary structural motifs; 5) availability of crystal data for the same protein system in multiple symmetry groups to explore the possible impact of different crystallographic contacts; and 6) availability of crystal data for the same protein system at multiple temperatures near room temperature, to probe the ability of the force field to capture the temperature dependence of the observables.

\subsubsection{Scorpion toxin}
\label{sub2:scorption_toxin}

Scorpion toxin (PDB ID 1AHO) is a 64-residue globular protein with a room-temperature X-ray data set at \qty{0.096}{\nano\meter} resolution \cite{smith_ab_1997}.
It has little regular secondary structure---one 9-residue helix and two short $\beta$-strands---but is stabilized by four disulfide bonds.
It was the subject of an early MD simulation \cite{cerutti_simulations_2010} that compared simulations of the crystal conducted with four protein force fields that would would now be considered obsolete.
The notable finding at the time was how diverse the simulation results were, even for simple metrics like average backbone structure and computed B-factors.
This study supported the idea that crystal simulations could be used for testing protein simulations and could report on both the stronger interactions that determine the conformation of an individual chain and also on the weaker (often solvent-mediated) interactions that stabilize the crystal lattice.
The small size of the unit cell was a more important consideration in 2010 than it would be today.

\subsubsection{Hen egg-white lysozyme}
\label{sub2:hewl}

Hen egg-white lysozyme (HEWL) was one of the first proteins whose structure was solved by X-ray crystallography, and it is widely used as an experimental model system, in part because of its ease of crystallization.
This protein has 129 residues, eight $\alpha$-helices, two $\beta$-sheets, and four disulfide bonds.
Given these constraints, simulations with different force fields are expected to reproduce the structural features of the deposited models with similar fidelity but to exhibit differences in fluctuations that are captured by observables such as electron densities and B-factors.

There are three RT crystal structures of the triclinic form (P1 space group) of HEWL (2LZT \cite{ramanadham_refinement_1990}, \qty{0.20}{\nano\meter} resolution; 6O2H \cite{meisburger_diffuse_2020}, \qty{0.12}{\nano\meter}; 4LZT \cite{walsh_refinement_1998}, \qty{0.095}{\nano\meter}), as well as a high resolution cryo version (2VB1 \cite{wang_triclinic_2007}, \qty{0.065}{\nano\meter}).
For all four structures, nitrate ions predominate in neutralizing the protein, and a large number of solvent waters are visible in the experimental electron density.
The crystal density of the solvent has been determined to high precision \cite{moreau_ice_2020} and is very close to that estimated by the MD simulations discussed below; this suggests that we know the amount of water in the unit cell to within an uncertainty of just a few water molecules. 

\begin{figure*}
    \centering
    \includegraphics[width=\linewidth]{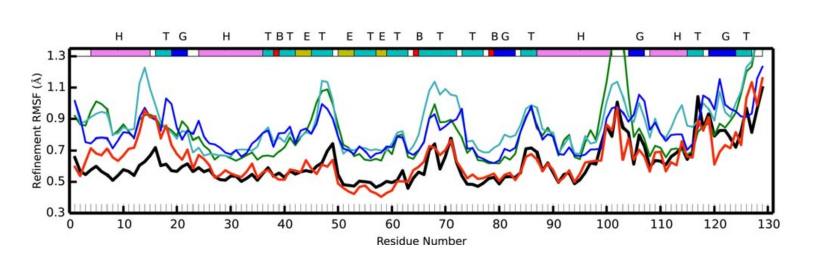}
    \caption{Refined C$\alpha$ root-mean-square fluctuations for four crystal simulations of HEWL, compared to experiment (black).
    Red: ff14SB, blue: CHARMM36; cyan: ff14ipq; green: ff99SB. 
    The colored band across the top describes the secondary structure (T: turn, E: b-sheet, H: a-helix, G:310 helix, B:isolated bridge).
    Adapted from Janowski et al. \cite{janowski_molecular_2016}.}
    \label{fig:hewl_xtal_rmsf}
\end{figure*}

Two studies of the triclinic HEWL crystal provide guidance for future studies.
First, Janowski et al. \cite{janowski_molecular_2016} simulated a triclinic supercell with 12 protein chains and made extensive comparisons to the reflection intensities from 4LZT.
Perhaps most insightful was a comparison of two newly refined atomic models, one refined against the experimental data and a second refined against the average electron density from a three-microsecond MD simulation.
The backbone of the structure refined against the simulated density was about \qty{0.04}{\nano\meter} away from the experimental structure, which is well outside the expected coordinate uncertainty of \qty{0.01}{\nano\meter} for a static crystal model, but small compared to differences between structural models of the same protein derived from crystal data in different space groups \cite{kovalevskiy_alphafold_2024}.
In addition, simulations using four protein force fields showed systematic differences in how close the B-factors refined against simulated densities were to experiment (Fig.~\ref{fig:hewl_xtal_rmsf}). Rather strikingly, B-factors computed with the ff14SB simulations are considerably closer to experiment (black) than those computed with three other protein force fields. This result further supports the utility of crystallographic data to benchmark protein force fields.

Second, Meisburger et al. \cite{meisburger_diffuse_2020} simulated triclinic HEWL with a primary focus on interpreting diffuse X-ray intensities.
Since collective motions of many protein chains are important for diffuse scattering, simulations were carried out using 1, 27, 125, and 343 unit cells, but they used only a single protein force field, so this study did not directly address the suitability of diffuse scattering for force field benchmarking.

There are also RT crystal structures for two other crystal forms---orthorhombic (1AKI \cite{artymiuk_structures_1982} and 8DYZ \cite{meisburger_robust_2023}) and tetragonal (8DZ7 \cite{meisburger_robust_2023})---of HEWL, and these may further probe the ability of force fields to capture crystal-packing interactions \cite{meisburger_robust_2023}.

\subsubsection{Crambin}
\label{sub2:crambin}

The small hydrophobic protein crambin, isolated from the seeds of the Abyssinian cabbage (\textit{Crambe abyssinica}), was found early on to form exceptionally well-ordered crystals \cite{teeter_water_1984} and has been used for the development of experimental phasing techniques.
Crambin has only 46 residues, contains both $\alpha$-helical and $\beta$-strand secondary structure, and is held together by disulfide bridges.
Of note here is a study combining RT X-ray and neutron diffraction to \qty{0.11}{\nano\meter} resolution (3U7T) and allowing for the modeling of anisotropic displacement parameters for select hydrogen atoms \cite{chen_room-temperature_2012}.
More recently, a new dataset of crambin was collected at RT to \qty{0.070}{\nano\meter} (9EWK), and the model refined to an R-factor of 0.0591 for reflections four standard deviations above background and of 0.0759 for all reflections, with stereochemical restraints released during the refinement process.
This model has one of the lowest R-factors in the PDB.
The estimated standard deviation for C-C bonds (a measure of the uncertainty in bond lengths for a model refined without restraints on bond lengths) was \qty{1.6}{\pico\meter} overall (\qty{0.6}{\pico\meter} for atoms with no alternate conformations).
These estimated standard deviations are similar in magnitude to those found in the Cambridge Structural Database (CSD) \cite{groom_cambridge_2016}, a database of small molecular crystals containing particularly reliable molecular geometries. 
NMR data has also been reported on recombinantly expressed crambin \cite{ahn_three-dimensional_2006}.

\subsubsection{Cyclophilin A}
\label{sub2:cypa}

Human cyclophilin A (CypA) is a proline \textit{cis}/\textit{trans} isomerase that is attractive for assessing force fields due to its well-studied conformational dynamics and high-resolution RT crystallographic data.
Several residues (Ser99, Phe113, Met61, Arg55) in the active site form a network that exhibits correlated conformational dynamics on a similar timescale (ms) as catalytic turnover, as measured by NMR relaxation experiments \cite{eisenmesser_intrinsic_2005}.
The alternate conformations in the CypA active site are primarily related to one another by transitions between side-chain rotamers, but they also involve backbone shifts that are well-modeled by the backrub motion \cite{davis_backrub_2006,friedland_simple_2008}.
More recently, multi-temperature crystallography across a series of eight temperatures added nuance to our understanding of CypA’s dynamic active-site network, revealing evidence for more complex, hierarchical coupling, in which dynamics for some active-site residues are dependent on the conformation of another key active-site residue (Phe113) \cite{keedy_mapping_2015}.
Diffuse scattering \cite{van_benschoten_measuring_2016} and X-ray solution scattering \cite{thompson_temperature-jump_2019,chen_solvent_2024} have also been measured for CypA.
Thus, CypA has rich dynamics that are well-studied by RT crystallography and other experiments and are ripe for comparison to MD simulations.

An attractive crystallography dataset for validating simulations of CypA is the high-resolution (\qty{0.120}{\nano\meter}) RT structure (PDB ID 4YUO) \cite{keedy_mapping_2015}, where various residues adopt two (A, B) or three (A, B, C) alternate conformations.
This complicates the creation of a single-conformer starting model for a given simulation; we recommend beginning with either state A or state B, which have distinct rotamers for Phe113, which is thought to be the linchpin of the network \cite{keedy_mapping_2015}.
One challenge in using CypA to validate simulations is that its catalysis and matching dynamics have been reported to be on the ms timescale \cite{eisenmesser_intrinsic_2005}.
However, multi-temperature crystallography also suggests motions at nanosecond timescales \cite{keedy_mapping_2015}.
Mutagenesis and solution scattering also point to a separate, independent loop region with faster dynamics \cite{thompson_temperature-jump_2019}.
Thus, even simulations at shorter than ms timescales are likely to uncover relevant dynamic features in CypA that can be useful for validation and force field optimization/development.
Moreover, CypA is moderately sized, with 165 amino acid residues, which enables reasonably long simulations.
Enhanced sampling techniques \cite{henin_enhanced_2022} that accelerate sampling of solute conformations such as replica exchange with solute tempering \cite{wang_replica_2011} may also be of use.
Time-scale issues are considered further in Section \ref{sub:best_practices_analysis}.
In addition to the \qty{0.120}{\nano\meter} RT crystal structure, a similarly high-resolution (\qty{0.125}{\nano\meter}) cryo structure of CypA (PDB ID 3K0M) \cite{fraser_hidden_2009} in the same crystal lattice is also available.

\subsubsection{Ubiquitin variants}
\label{sub2:ubiquitin}

Computational protein design and directed evolution have been combined to generate ubiquitin variants with enhanced affinity to the protein USP7 through stabilization of a specific conformation of the $\beta$1$\beta$2 loop \cite{zhang_conformational_2013}.
In a subsequent study, RT X-ray crystallography was used to examine the structural basis for the increasing affinity of ubiquitin variants over the course of the design/selection process \cite{biel_flexibility_2017}.
The resulting high-resolution RT crystal structures revealed that the earlier "core" mutant of ubiquitin (with 6 mutations) exhibits multiple discrete alternate conformations of the $\beta$1$\beta$2 loop, whereas the later "affinity-matured" mutant (with 3 additional mutations) adopts a singular conformation of this loop.
Also, elsewhere in the protein, both variants exhibit alternate conformations for a peptide flip of residues 52-53.

High-resolution RT crystallography data and models are available for both ubiquitin mutant proteins, at resolutions of \qty{0.112}{\nano\meter} for the "core" mutant (PDB ID 5TOF) and \qty{0.108}{\nano\meter} for the "affinity-matured" mutant (PDB ID 5TOG).
Simulations based on these datasets should ideally capture the differences in $\beta$1$\beta$2 loop conformational heterogeneity observed in the crystallographic electron density maps for the two variants (more flexible for the core mutant, more rigid for the affinity-matured mutant) as well as the peptide flip shared by both variants.
For this system, a number of metrics could be used to quantify the match between simulations and experimental data: root-mean-square fluctuations (RMSF) of backbone atoms in the loop for simulations vs. for multi-conformer crystal structures, local real-space fit to the electron density map in the $\beta$1$\beta$2 loop region, recapitulation of fractional occupancies for the different loop conformations (perhaps after clustering the simulation snapshots), etc.
Thus, the high-resolution RT crystallography datasets available for these ubiquitin variants provide an opportunity to benchmark force fields for their ability to provide accurate simulations of protein backbone conformational heterogeneity, as well as accurate predictions of the effects of amino acid substitutions on conformational heterogeneity.

\subsubsection{Protein tyrosine phosphatase 1B}
\label{sub2:ptp1b}

Human protein tyrosine phosphatase 1B (PTP1B; also known as PTPN1) exhibits structural dynamics across a range of time scales and length scales, as revealed by numerous structural biophysics experiments.
Recently, multi-temperature X-ray crystallography of apo PTP1B at temperatures from cryo to RT \cite{keedy_expanded_2018} provided insights into correlated conformational heterogeneity.
The resulting series of crystal structures featured alternative conformations, each modeled with partial occupancy, for the active-site WPD loop (open vs. closed) as well as distal allosteric regions.
As temperature increased, the WPD loop shifted crystallographic occupancy from the closed to the open state.
Simultaneously, the distal $\alpha$7 helix, a key component of PTP1B’s allosteric network, shifted occupancy from the ordered state to a disordered state, moving into an adjacent solvent channel in the crystal lattice.
Several residues between these regions exhibited smaller-scale conformational shifts between alternate conformations of the WPD loop and $\alpha$7 helix, mimicking the shifts seen previously for an allosteric inhibitor that displaces $\alpha$7 \cite{wiesmann_allosteric_2004}.

Other studies have used NMR spectroscopy to characterize timescales of motion for various regions of PTP1B.
NMR relaxation experiments showed that the active-site WPD loop closes on a timescale corresponding to the rate of catalysis (ms) \cite{whittier_conformational_2013}.
Beyond the active site, NMR relaxation experiments, mutagenesis, and molecular dynamics simulations restrained by NMR chemical shifts showed that faster dynamics are key to allosteric regulation via $\alpha$7 \cite{choy_conformational_2017}.
Thus, PTP1B exhibits motions that may be amenable to various different types of simulations, from short, traditional simulations to long, enhanced-sampling simulations.

Both of the major states of the protein are modeled in the \qty{0.174}{\nano\meter} RT (278 K) crystal structure of apo PTP1B (PDB ID 6B8X): the closed state (alternate conformation A) and the open state (B).
Most regions of the structure are modeled with either no alternate conformations, or both A and B conformations.
By contrast, because it is disordered in the open state of the protein, the $\alpha$7 helix is modeled as only the A conformation with partial occupancy, with no coordinates for the B conformation.
Note that crystallography was performed with residues 1-321 of PTP1B, but only residues 1-298 are visible in the electron density, even in the closed state with $\alpha$7 ordered; the remaining residues are always disordered in an open region of bulk solvent within the crystal lattice.
The wealth of types and extents of conformational heterogeneity it features make this a promising candidate for force field evaluation.
Simulations of PTP1B based on 6B8X should be initiated from either the open state or the closed state of the protein and should be assessed on the basis of their ability to recapitulate the allosteric coupling observed in various experiments: as the WPD loop opens, the allosteric network should shift toward the corresponding open-like state, and $\alpha$7 should become disordered.
In addition to unbiased simulations \cite{yeh_conserved_2023}, one could perform biased or guided simulations to enforce a shift in one region (WPD or $\alpha$7), then examine whether the other region allosterically responds as expected \cite{crean_sequence_2024}.

\subsubsection{Endoglucanase}
\label{sub2:endoglucanase}

Combined X-ray and neutron-diffraction data are available for endoglucanase (EG) from \textit{Phanerochaete chrysosporium} at room temperature (PDB 3X2P) \cite{nakamura_newtons_2015}.
As noted in Section \ref{sub2:neutron}, the neutron data make it possible to use experimentally derived protonation states for protein residues \cite{blakeley_neutron_2009,chen_direct_2012,ashkar_neutron_2018}, removing a potential cause of modeling errors and uncertainties, while also allowing direct determination of the orientation of ordered waters and other factors.
Water structure in this case is also much more clearly defined, and amenable to careful curation \cite{blakeley_15-k_2004,wall_biomolecular_2019}.
Note, however, that this structure includes a non-standard amino acid and an oligosaccharide ligand.

In one simulation study \cite{wall_biomolecular_2019}, the EG-cellopentaose complex from 3X2P was re-refined from \qty{0.15}{\nano\meter} neutron and \qty{0.10}{\nano\meter} X-ray diffraction data, with careful attention to placing H/D atoms based on neutron scattering data and H-bond interactions with the local environment.
The re-refined structure featured several differences in protonation states with respect to the deposited model and the replacement of an imidic acid form for an asparagine sidechain in the deposited model with canonical asparagine.
This particular study focused on the ability of MD simulations to correctly position water molecules defined by the crystallographic studies.
This was accomplished by computing electron density maps from crystalline simulations of a 2 $\times$ 2 $\times$ 2 periodic supercell under several different solvent conditions, with several different restraints of the protein heavy atoms and ligand atoms.
Clear indications of force field limitations were identified.
For example, although recovery of experimentally observed crystallographic waters was over 90\% when the protein was restrained modestly to crystallographic positions, this fraction dropped to 50\% without restraints.
The drop traced to small, local protein motions that disrupted entire water networks \cite{wall_biomolecular_2019}, consistent with prior work \cite{lexa_full_2011}.

While this study focused primarily on recovery of crystallographic water molecules, it suggests a route forward in terms of benchmarking force fields on this and similar systems \cite{wall_biomolecular_2019}.
The carefully curated re-refined structures provide a valuable starting point and should be used as input for simulations.
Force fields could be benchmarked by repeating the simulation approach used here, while comparing results at different restraint strengths.
Presumably, as force fields are improved, this will result in better recovery of crystallographic water structures at lower protein restraint strengths or with no restraints.
Additional benchmarking studies could more closely focus on how well various force fields preserve the structure of the protein.

\subsubsection{Staphylococcal nuclease}
\label{sub2:staph_nuclease}

Staphylococcal nuclease (SNase) has provided a particularly valuable dataset for diffuse scattering studies \cite{wall_three-dimensional_1997}.
Until fairly recently, this was the only complete, high-quality, three-dimensional (i.e. sampled on a 3-D reciprocal space grid using data from many diffraction images obtained at different crystal orientations) diffuse scattering dataset available for a protein crystal \cite{wall_conformational_2014}.
Crystalline MD simulations greater than \qty{1}{\micro\second} in length allowed calculation of diffuse scattering intensity for direct comparison with experiment \cite{wall_conformational_2014}.
More recent work \cite{wall_internal_2018} extended these simulations to cover a 2 $\times$ 2 $\times$ 2 supercell, with roughly \qty{5}{\micro\second} of data, and obtained improved agreement with experimental scattering data, perhaps because the simulations used a starting structure that modeled missing terminal residues and a bound ligand that better represented the system in the crystal experiment.
Different force fields gave different levels of agreement with experimental diffuse scattering patterns \cite{wych_liquid-like_2019}, supporting the concept of using such data to benchmark protein force fields.

\section{Best practices for setup and analysis of benchmark simulations}
\label{sec:best_practices}

\subsection{Setup of benchmark simulations}
\label{sub:best_practices_setup}

Simulations intended to benchmark protein force fields against experimental observables should strive to replicate the conditions under which those observables were measured as closely as possible. Below we provide specific recommendations to achieve this goal.

Benchmark simulations must be started from a set of initial coordinates for the protein.
For folded proteins with structural models refined against experimental data, the initial coordinates can simply be those of the appropriate structural model.
When both crystal and NMR models are available, one should select the model that corresponds to the desired conditions of the simulation, i.e. NMR models should be selected for simulations of proteins in dilute solution while crystal models should be selected for simulations of protein crystals.
For short peptides, the initial coordinates are often set to an extended conformation in which all backbone dihedrals are set to \ang{180}.
For disordered proteins, a structural model for a partially folded conformation can be used when available, and an extended conformation can be used for residues not included in the structural model.
An extended conformation can also be used for the entire disordered protein, although this setup requires a very large simulation box with many solvent molecules.
To avoid large boxes, several methods \cite{feldman_fast_2000,feldman_probabilistic_2002,ozenne_flexible-meccano_2012,estana_realistic_2019,ferrie_unified_2020,teixeira_idpconformergenerator_2022} can be used to generate one or more initial conformations consistent with an expected radius of gyration.

While it is common practice to remove hydrogen atoms present in structural models and add them back using standard residue libraries, it is essential to assign correct protonation states to titratable sidechains and protein termini.
High resolution crystal models, especially those refined against neutron diffraction data, can include the coordinates of hydrogen atoms and thus give high confidence in protonation states.
Otherwise, the choice of protonation states should generally reflect the expected protonation based on the pH of the experiments used to measure the observables.
However, the local environment of a residue---such as hydrogen bonds to nearby residues, long-range electrostatics, and solvation---may indicate a protonation state different from the expected one at the experimental pH.
Similarly, the local environment of cysteine residues can indicate whether they should be modeled with a disulfide bond.
Tools such as PROPKA \cite{olsson_propka3_2011} and PDB2PQR \cite{jurrus_improvements_2018} can assign protonation states and disulfide bonds based on local environment and a specified pH.

Most benchmark simulations model water molecules explicitly, and the water model should be considered part of the force field being benchmarked just as much as the parameters applied to the protein.
Indeed, the choice of force field parameters for the water can significantly impact the resulting protein ensembles \cite{piana_water_2015,tian_ff19sb_2020,coppa_accelerated_2023}.
In particular, the widely used 3-point water model TIP3P \cite{jorgensen_comparison_1983} favors compact structures excessively \cite{tian_ff19sb_2020,coppa_accelerated_2023} and has low viscosity compared to experimental water \cite{kadaoluwa_pathirannahalage_systematic_2021}.
The consequences will be more pronounced for simulations of disordered proteins and for calculations of kinetic observables, such as NMR spin relaxation rates.
In such applications, simulations using TIP3P should be expected to exhibit some disagreement with experiment due to the properties of the water model rather than the protein force field.

For crystal simulations, the simplest setup is a periodic box containing a single unit cell from the crystal.
A single unit cell might be adequate for some Bragg data applications, or for modeling the isotropic component of diffuse scattering data. 
Other applications might require a supercell in which the periodic simulation box includes multiple copies of the crystal unit cell that can move independently.
For example, in diffuse scattering studies, using two copies of the unit cell along each lattice vector avoids artificial correlations between pairs of atoms that are separated by a distance smaller than the unit cell, and can substantially increase the agreement with the anisotropic component of diffuse scattering data \cite{wall_internal_2018}.
An appropriate choice of the supercell also is required for direct comparison of simulations with the full set of diffuse scattering measurements: in general, a full dataset with $N_h$, $N_k$, and $N_l$ points sampled per reciprocal lattice vector along Miller indices $hkl$ can be simulated using a box with $N_h$, $N_k$, and $N_l$ copies of the unit cell along lattice vectors $\vec{a}$, $\vec{b}$, and $\vec{c}$.
Representative lattices generated by this approach may be, e.g., 2$\times$2$\times$2 or 7$\times$7$\times$5.

For dilute solution simulations, the periodic box should contain the protein and a large enough number of solvent molecules to effectively screen interactions between periodic images of the protein (see next paragraph).
Many system-building tools provide automated methods to construct periodic boxes filled with pre-equilibrated solvent around a solute with the box geometry defined by user-provided shape and dimensions.
While triclinic simulation boxes are straightforward to build and visualize, other polyhedra that can fill space using only translations can provide the same minimum distances between periodic images with lower volume and thus faster simulation speed.
The rhombic dodecahedron \cite{wang_superposition_1972} has the smallest volume of these space-filling polyhedra and is the recommended box shape for simulations of solutes that are roughly spherical, including most globular proteins.

The size of the periodic simulation box can be specified using either the total box length or the solvent padding, i.e., the shortest acceptable distance between a protein atom and the edge of the simulation box.
If the solvent padding is shorter than \qty{1}{\nano\meter}, enough space for two to three shells of water, then interactions between periodic images can perturb observed quantities such as solvation free energies and secondary structure preferences 
\cite{mehra_cell_2019,gapsys_importance_2020}.
Based on these considerations, for folded proteins that are expected to sample only compact structures during the simulation, a solvent padding of \qtyrange{1.0}{1.5}{\nano\meter} is sufficient to avoid such artifacts.
For disordered proteins, a larger solvent padding should be used to accommodate these less compact structures.
A common practice for disordered proteins is to use a total box length that is twice the radius of gyration plus \qty{2}{\nano\meter}.

For proteins with a net charge, counterions should be added to the simulation box to neutralize the system.
If the observables targeted by the simulation were measured in a solution containing salt, then additional ion pairs should be added to the simulation box to model the experimental ionic strength of the bulk salt.
If the protein has no net charge, the desired number of ion pairs can be estimated from the number of water molecules used to solvate the system $N_{\mathsf{water}}$.

\begin{equation}
\label{eqn:ions_neutral}
N_{\mathsf{ions}} = \frac {N_{\mathsf{water}}} {C_{\mathsf{water}}} C_{\mathsf{ions}}
\end{equation}

\noindent where $C_{\mathsf{water}}$ is the expected concentration of bulk water in dilute solution, \qty{55.4}{\mol\per\liter}, and $C_{\mathsf{ions}}$ is experimental concentration of bulk salt.
If the protein has a net charge, then the solvent near the protein will be depleted of ions of like charge and enriched in ions of opposite charge, relative to bulk \cite{schmit_sltcap_2018}.
In this case, the desired number of ion pairs can be estimated using the SLTCAP method (this expression is equivalent to Eq. 5 in Schmit et al.) \cite{schmit_sltcap_2018}.

\begin{equation}
\label{eqn:ions_charged}
\begin{gathered}
N_{\mathsf{ions}} = \frac {N_{\mathsf{water}}} {C_{\mathsf{water}}} C_{\mathsf{ions}} \left( \sqrt{1 + \lambda^2} - \lambda \right) \\
\lambda = \frac {\left| Q \right| C_{\mathsf{water}}} {2 N_{\mathsf{water}} C_{\mathsf{ions}}}
\end{gathered}
\end{equation}

In addition to salt, co-solutes such as buffering agents or crowding agents may be present in solutions used to measure observables.
Such co-solutes are particularly important for protein crystals, which often require co-solutes in the mother liquor in order to induce crystallization.
It is sometimes necessary to model co-solutes explicitly to accurately describe the solvent environment within protein crystals \cite{wall_biomolecular_2019}.
Modeling co-solutes explicitly can improve the fidelity of benchmark simulations, but care must be taken to ensure that appropriate force field parameters are available for the co-solutes and that sampling is long enough to observe changes in cosolute locations that can influence protein conformations and convergence of the protein ensemble.

Because the observables of interest here are typically measured at constant pressure and temperature, benchmark simulations should sample from the isobaric-isothermal (NPT) ensemble.
Some commonly used methods for controlling the pressure and temperature in simulations produce samples from different, unphysical ensembles.
In particular, Berendsen or weak-coupling methods fail to preserve the equipartition of energy and can lead to artifacts such as the flying ice cube effect \cite{harvey_flying_1998,braun_anomalous_2018}.
These methods should be avoided in favor of methods that sample correctly from physical ensembles such as Langevin dynamics, the canonical velocity rescaling thermostat, and the Monte Carlo barostat.
Additional care is needed when computing kinetic quantities, as some thermostats---most notably Langevin dynamics, with its added frictional damping and random forces---may introduce significant kinetic artifacts.

\subsection{Analysis of benchmark simulations}
\label{sub:best_practices_analysis}

It is recommended to discard the initial segment of a molecular dynamics simulation before collecting the frames that will be used to calculate observables.
The discarded initial segment, often termed the equilibration or burn-in run, allows the system to relax from its initial configuration, which may be far from the equilibrium distribution dictated by the force field.
Several methods are available to set the length of the equilibration run. 
One common practice is to monitor the volume or density of the barostatted system and to discard frames until this quantity reaches a plateau.
An equilibration time can also be chosen based on the autocorrelation function of a time series of an observable \cite{chodera_simple_2016}.

Precise estimation of quantities from molecular dynamics simulations requires a large number of samples drawn from the system's ensemble.
It is good practice to assess the convergence of simulations by determining whether estimates of observables are robust to changes in the set of conformational samples used to estimate them.
A number of methods have been developed to assess convergence \cite{grossfield_best_2019}.
In one simple approach, the time series of observable estimates is visualized to check for drift over the course of the simulation.
Blocking analysis, a more rigorous approach, divides the trajectory into blocks and looks for convergence of the standard error of the mean as the block size is increased \cite{flyvbjerg_error_1989}.

Assessment of convergence will identify when the populations of states visited during a simulation are not known precisely, but such analysis cannot identify when a simulation fails entirely to visit regions of a protein's configuration space that carry significant weight in the Boltzmann distribution.
This failure mode can occur, for example, when high free energy barriers exist between minima that contribute to the protein's ensemble in the experimental measurements.
Diagnosing this problem requires some knowledge of the proteins being simulated, either from experimental measurements or from a reference ensemble known to model the system well.
If benchmark simulations of a particular protein consistently fail to visit relevant regions of the protein's configuration space across multiple force fields, enhanced sampling methods \cite{henin_enhanced_2022} can accelerate barrier crossing, though at the cost of additional complexity in obtaining estimates of observables and uncertainties \cite{grossfield_best_2019}.

When comparing sets of simulations against experimental data, it is essential to determine whether the differences are statistically significant \cite{van_gunsteren_validation_2018}.
Statistical uncertainties in simulated data are perhaps most reliable if obtained by running multiple independent simulations but can also be obtained by analysis of a single simulation \cite{grossfield_best_2019}.
The raw experimental data also have uncertainties, and additional uncertainty may result from model assumptions, such as the values of the Karplus parameters for $^3J$-values or the representation of NOESY intensities as upper bounds to interatomic distances.
When multiple proteins are simulated and several properties are considered, it may not be straightforward to determine if two sets of simulations are significantly different or even which simulation shows the overall better agreement with the available experimental data \cite{wassenaar_effect_2006,villa_how_2007,stroet_validation_2024}.
The differences in a particular property between different force fields will be affected by both the variability due to the choice of protein and the variability between independent replicate simulations.
A statistical approach that takes mixed effects explicitly into account \cite{stroet_validation_2024} may be most appropriate.

\section*{Author Contributions}

All authors contributed to conceptualizing and defining the scope of the review article.
Contributions to the initial draft of the manuscript were as follows:
\begin{itemize}
    \item Section \ref{sec:overall} on recommendations by C.E.C. and M.K.G.
    \item Section \ref{sub2:chem_shift} on chemical shifts by L.T.C. and P.J.R.
    \item Section \ref{sub2:j_coupling} on $J$ couplings by P.J.R.
    \item Section \ref{sub2:rdc} on RDCs by K.L.-L.
    \item Section \ref{sub2:noesy} on NOESY by C.E.C. and C.O.
    \item Section \ref{sub2:spin_relax} on spin relaxation by O.H.S.O.
    \item Section \ref{sub2:pre} on PREs by K.L.-L.
    \item Section \ref{sub2:beauchamp} on Beauchamp dataset by P.J.R.
    \item Section \ref{sub2:designed_beta} on designed peptides by V.A.V.
    \item Section \ref{sub2:stroet} on Stroet dataset by C.O.
    \item Section \ref{sub2:mao} on Mao dataset by P.J.R.
    \item Section \ref{sub2:robustelli} on Robustelli dataset by P.J.R.
    \item Section \ref{sub2:spin_relax_datasets} on spin relaxation datasets by O.H.S.O.
    \item Section \ref{sub2:salt_bridge} on salt bridge stability dataset by L.T.C.
    \item Section \ref{sub2:bragg} on Bragg diffraction by D.A.C.
    \item Section \ref{sub2:b_factors} on B-factors by J.C.-H.C.
    \item Section \ref{sub2:alt_confs} on alternate conformations by D.A.K.
    \item Section \ref{sub2:diffuse} on diffuse scattering by M.E.W., D.C.W. and D.A.C.
    \item Section \ref{sub2:neutron} on neutron diffraction by J.C.-H.C.
    \item Section \ref{sub2:scorption_toxin} on scorpion toxin by D.A.C.
    \item Section \ref{sub2:hewl} on lysozyme by D.A.C. and J.C.-H.C.
    \item Section \ref{sub2:crambin} on crambin by J.C.-H.C.
    \item Section \ref{sub2:cypa} on cyclophilin A by D.A.K.
    \item Section \ref{sub2:ubiquitin} on ubiquitin by D.A.K.
    \item Section \ref{sub2:ptp1b} on PTP1B by D.A.K.
    \item Section \ref{sub2:endoglucanase} on endoglucanase by D.L.M.
    \item Section \ref{sub2:staph_nuclease} on SNase by D.L.M. and M.E.W.
    \item Section \ref{sec:best_practices} on best practices by C.E.C. and C.O.
\end{itemize}
All authors contributed to revisions of the initial draft.
C.E.C. and M.K.G. coordinated and managed the writing and revision of the manuscript.

For a more detailed description of author contributions, see the GitHub issue tracking and changelog at \githubrepository.

\section*{Other Contributions}

The authors acknowledge Bernard Brooks, David Cerutti, Thomas Cheamtham III, John Chodera, Gabriel Rocklin, Adrian Roitberg, Benoit Roux, Michael Shirts, and Junmei Wang for early discussions about benchmarking datasets.

For a more detailed description of contributions from the community and others, see the GitHub issue tracking and changelog at \githubrepository.

\section*{Potentially Conflicting Interests}

L.T.C. serves on the scientific advisory board of OpenEye Scientific Software and is an Open Science Fellow with Psivant Sciences.
M.K.G. has an equity interest in and is a cofounder and scientific advisor of VeraChem LLC and a member of the Scientific Advisory Board of InCerebro Co., Ltd.
D.L.M. serves on the scientific advisory boards of OpenEye Scientific Software and Anagenex, and is an Open Science Fellow with Psivant Sciences.
M.E.W. is a consultant for Eli Lilly and Company.
K.L.-L. holds stock options in and is a consultant for Peptone Ltd.

\section*{Funding Information}

C.E.C. acknowledges funding from the National Institutes of Health (R01GM132386 and F32GM150240) and the
Open Force Field Consortium.
L.T.C. acknowledges support from the National Science Foundation (CHE-1807301)
D.A.K. acknowledges funding from the National Institutes of Health (R35GM133769).
These findings are solely of the authors and do not necessarily represent the views of the NIH,
the NSF, or any other funder.
O.H.S.O. acknowledges the Academy of Finland (grants 315596 and 319902) for financial
support.
P.J.R. acknowledges support from the National Institutes of Health under award R35GM142750.
V.A.V. acknowledges funding from the National Institutes of Health (R01GM123296).
D.C.W. and M.E.W. acknowledge support from the Exascale Computing Project (grant No.
17-SC-20-SC), a collaborative effort of the US Department of Energy Office of Science and
National Nuclear Security Administration.
D.C.W. acknowledges support from the Center for Nonlinear Studies at Los Alamos National
Laboratory (under US Department of Energy contract No. 89233218CNA000001 to Triad
National Security).
M.K.G. acknowledges funding from the National Institutes of Health (R01GM132386).
Co-funded by the European Union (ERC, DynaPLIX, SyG-2022 101071843, to K.L.-L.).
Views and opinions expressed are however those of the authors only and do not necessarily reflect those of the European Union or the European Research Council.
Neither the European Union nor the granting authority can be held responsible for them.

\section*{Author Information}
\makeorcid

\bibliography{review-protein-benchmark-datasets}


\appendix
\section{Effective distances probed in NOESY experiments}
\label{app:noesy_distances}

The NOE between two nuclei, $A$ and $B$, depends primarily on their dipolar cross relaxation rate, $\sigma_{AB}$.

\begin{equation}
\label{eqn:noe_cross_relaxation}
\sigma_{AB} = K \left( 3 J(\omega_A + \omega_B) - \frac {1} {2} J(\omega_A - \omega_B) \right)
\end{equation}

\noindent where

\begin{equation}
\label{eqn:noe_constant}
K = \left( \frac {\mu_0} {4 \pi} \right)^2 \frac {\hbar^2 \gamma_A^2 \gamma_B^2} {10}
\end{equation}

\noindent and $\mu_0$ is the vacuum permeability, $\hbar$ is the reduced Planck constant, $\gamma_A$ and $\gamma_B$ are the gyromagnetic ratios of the nuclei, and $\omega_A$ and $\omega_B$ are the Larmor frequencies of the nuclei.
$J(\omega)$ is a spectral density function associated with the relative geometry of the two spins, whose correlation function is determined by the dynamics of the lattice. It essentially describes the spectrum of the power available from the lattice to activate the dipolar cross relaxation pathway.
Thus, the spectral density function is given by the Fourier transformation of the correlation function for the internuclear vector, $C(t)$:
\begin{equation}
\label{eqn:noe_spectral_density}
\begin{gathered}
J(\omega) = \int_0^\infty \mathsf{d}t C(t) \cos (\omega t) \\
C(t) = \left \langle \frac {3 \cos^2 \theta_{AB}(t) - 1} {2 R_{AB}(0)^3 R_{AB}(t)^3} \right \rangle
\end{gathered}
\end{equation}

\noindent where $R_{AB}(t)$ is the distance between the nuclei at time $t$ and $\theta_{AB}(t)$ is the angle between the internuclear vectors at time $0$ and time $t$.

As discussed in Section~\ref{sub2:noesy}, NOESY experiments provide effective mean distances between nuclei.
We next discuss two common approaches to determining effective distances, and the conditions under which they apply.
For a more detailed treatment, see Neuhaus \& Williamson \cite{neuhaus_nuclear_2000} and Vogeli \cite{vogeli_nuclear_2014}.
For simplicity, we consider the case of a  homonuclear NOESY experiment, where---to good approximation---$\omega_A = \omega_B = \omega$, so the dipolar cross relaxation rate reduces to

\begin{equation}
\label{eqn:noe_homonuclear}
\sigma_{AB} = K \left[ 3\: J(2 \omega) - \frac {1} {2} J(0) \right]
\end{equation}

For a rigid molecule that tumbles isotropically with a rotational diffusion time constant $\tau_C$, the spectral density function is given by

\begin{equation}
\label{eqn:noe_rigid_spectral_density}
J(\omega) = \frac {1} {R_{AB}^6} \frac {2 \tau_C} {1 + \omega^2 \tau_C^2}
\end{equation}

\noindent where $R_{AB}$ is the constant distance between the nuclei. In this case, the dipolar cross relaxation rate becomes

\begin{equation}
\label{eqn:noe_rigid_cross_relaxation}
\sigma_{AB} = \frac {K} {R_{AB}^6} \left( \frac {6 \tau_C} {1 + 4 \omega^2 \tau_C^2} - \tau_C \right)
\end{equation}

For a non-rigid molecule, it is often assumed that internal motions, i.e. intramolecular motions, are uncorrelated with molecular tumbling and can be characterized by a single exponential decay with a time constant $\tau_{\mathsf{int}}$.
Under these assumptions, called the Lipari-Szabo or model-free approximation, the spectral density is given by

\begin{equation}
\label{eqn:ls_spectral_density}
J(\omega) = \left \langle \frac {1} {R_{AB}^6} \right \rangle \left( S^2 \frac {2 \tau_C} {1 + \omega^2 \tau_C^2} + \left( 1 - S^2 \right) \frac {2 \tau_{\mathsf{tot}}} {1 + \omega^2 \tau_{\mathsf{tot}}^2} \right)
\end{equation}

\noindent where

\begin{equation}
\label{eqn:ls_total_tumbling}
\frac {1} {\tau_{\mathsf{tot}}} = \frac {1} {\tau_C} + \frac {1} {\tau_{\mathsf{int}}}
\end{equation}

\noindent and

\begin{equation}
\label{eqn:ls_order_parameter}
S^2 = \left \langle \frac {1} {R_{AB}^6} \right \rangle^{-1} \frac {1} {5} \sum_{m=-2}^2 \left \langle \frac {Y_{2m}(\theta_{AB}, \phi_{AB})} {R_{AB}^3} \right \rangle^2
\end{equation}

\noindent Here, $Y_{2m}(\theta_{AB}, \phi_{AB})$ are the second order spherical harmonics defined using the spherical coordinates of the internuclear displacement in the molecular frame, and $S^2$ is an order parameter that is determined by the plateau value of the correlation function $C(t)$ at long times and characterizes the orientational freedom of the internuclear displacement.
The order parameter varies between one---when the internuclear displacement is rigid in the molecular frame---and zero---when the internuclear displacement tumbles isotropically in the molecular frame.
In the Lipari-Szabo model, the dipolar cross relaxation rate becomes

\begin{equation}
\label{eqn:ls_cross_relaxation}
\begin{gathered}
\sigma_{AB} = K \left \langle \frac {1} {R_{AB}^6} \right \rangle \left[ S^2 \left( \frac {6 \tau_C} {1 + 4 \omega^2 \tau_C^2} - \tau_C \right) \right. \\
\left. + (1 - S^2) \left( \frac {6 \tau_{\mathsf{tot}}} {1 + 4 \omega^2 \tau_{\mathsf{tot}}^2} - \tau_{\mathsf{tot}} \right) \right]
\end{gathered}
\end{equation}

This expression is difficult to use in practice because the time constant for internal motions, and thus $\tau_{\mathsf{tot}}$, cannot be easily measured.
Fortunately, we can simplify this expression by considering limiting cases when internal motions are either much slower or much faster than molecular tumbling.
In the limit where internal motions are much slower than molecular tumbling, i.e. $\tau_{\mathsf{int}} \gg \tau_C$, $\tau_{\mathsf{tot}} \approx \tau_C$ such that the terms containing the order parameter $S^2$ cancel and the dipolar cross relaxation rate loses its dependence on the angular degrees of freedom.

\begin{equation}
\label{eqn:ls_slow_cross_relaxation}
\sigma_{AB} = K \left \langle \frac {1} {R_{AB}^6} \right \rangle \left( \frac {6 \tau_C} {1 + 4 \omega^2 \tau_C^2} - \tau_C \right)
\end{equation}

\noindent This expression looks like the expression for a rigid molecule (Eq. \ref{eqn:noe_rigid_cross_relaxation}) with the internuclear distance probed by the NOE enhancement replaced by the following effective distance

\begin{equation}
\label{eqn:ls_slow_effective_distance}
R_{AB}^{\mathsf{slow}} = \left \langle \frac {1} {R_{AB}^6} \right \rangle^{-1/6}
\end{equation}

\noindent Thus, in the limit when internal motions are much slower than molecular tumbling, the NOE averages the variation in internuclear distance due to internal motions raised to the reciprocal sixth power and is independent of the orientation of the internuclear displacement.

In the opposite limit, i.e.,  when internal motions are much faster than molecular tumbling ($\tau_{\mathsf{int}} \ll \tau_C$), the term in Equation~\ref{eqn:ls_spectral_density} containing $(1 - S^2)$ becomes negligible, so

\begin{equation}
\label{eqn:ls_fast_cross_relaxation}
\sigma_{AB} = K \left \langle \frac {1} {R_{AB}^6} \right \rangle S^2 \left( \frac {6 \tau_C} {1 + 4 \omega^2 \tau_C^2} - \tau_C \right)
\end{equation}

\noindent In this case, the effective distance probed by the NOE enhancement is

\begin{equation}
\label{eqn:ls_fast_effective_distance_angular}
R_{AB}^{\mathsf{fast,ang}} = \left( \frac {1} {5} \sum_{m=-2}^2 \left \langle \frac {Y_{2m}(\theta_{AB}, \phi_{AB})} {R_{AB}^3} \right \rangle^2 \right)^{-1/6}
\end{equation}

\noindent The effective distance can be simplified further by assuming that the distances $R_{AB}$ exhibit greater variations than the angles in the spherical harmonic functions.
Under this assumption, the angular dependence is ignored, and the effective distance is chosen to be

\begin{equation}
\label{eqn:ls_fast_effective_distance}
R_{AB}^{\mathsf{fast}} = \left( \left \langle \frac {1} {R_{AB}^3} \right \rangle^2 \right)^{-1/6} = \left \langle \frac {1} {R_{AB}^3} \right \rangle^{-1/3} \leq R_{AB}^{\mathsf{fast,ang}}
\end{equation}

\noindent where the inequality comes from the observation that the spherical harmonic functions are normalized such that the angular dependence in Eq. \ref{eqn:ls_fast_effective_distance_angular} must reduce the magnitude of quantity averaged over and thus increase the effective distance.

\section{The Redfield equations for NMR spin relaxation}
\label{app:spin_relax_redfield}

Spin relaxation rates in proteins are often measured for spins coupled through a covalent bond, and a heteronuclear NOE can also be measured for the same bond.
The Redfield equations \cite{redfield_theory_1965} describe how the spin relaxation rates and NOE depend on the dynamics of the bond within a molecule.
For the backbone $^{15}$N amide bond in proteins, the Redfield equations are

\begin{equation}
\label{eqn:redfield}
\begin{gathered}
R_1 = \frac {1} {T_1} = \frac {K} {R_{NH}^6} \left( 3 J(\omega_H + \omega_N) + \frac {3} {2} J(\omega_N) + \frac {1} {2} J(\omega_H - \omega_N) \right) \\
+ \frac {\left( \omega_N \Delta \sigma \right)^2} {15} J(\omega_N) \\
R_2 = \frac {1} {T_2} = \frac {1} {2} \frac {K} {R_{NH}^6} \left( 3 J(\omega_H + \omega_N) + \frac {3} {2} J(\omega_N) + \frac {1} {2} J(\omega_H - \omega_N) \right. \\
\left. + 3 J(\omega_H) + 2 J(0) \right) + \frac {\left( \omega_N \Delta \sigma \right)^2} {90} \left( 3 J(\omega_N) + 4 J(0) \right) \\
\mathsf{NOE} = \frac {K} {R_{NH}^6} \left( 3 J(\omega_H + \omega_N) + \frac {1} {2} J(\omega_H - \omega_N) \right) \times \frac {\gamma_H} {\gamma_N} T_1
\end{gathered}
\end{equation}

\noindent where, similar to Appendix \ref{app:noesy_distances},

\begin{equation}
\label{eqn:spin_relaxation_constant}
K = \left( \frac {\mu_0} {4 \pi} \right)^2 \frac {\hbar^2 \gamma_N^2 \gamma_H^2} {10}
\end{equation}

\noindent and $\mu_0$ is the vacuum permeability, $\hbar$ is the reduced Planck constant, $\gamma_N$ and $\gamma_H$ are the gyromagnetic ratios of the nuclei, $\omega_N$ and $\omega_H$ are the Larmor frequencies of the nuclei, $R_{NH}$ is the amide bond length (assumed constant), and $J(\omega)$ is the spectral density function (Eq. \ref{eqn:noe_spectral_density}).
$\Delta \sigma$ is the chemical shift anisotropy for the $^{15}$N spin, usually taken to be 160 ppm.

The length of the N-H bond is typically assumed to be constant, so that the spectral density function is given by the Fourier transformation of the rotational correlation function

\begin{equation}
\label{eqn:spin_relaxation_correlation_function}
C_{\mathsf{rot}}(t) = \left \langle \frac {3} {2} \cos \theta_{AB}(t) - \frac {1} {2} \right \rangle
\end{equation}

Under the Lipari-Szabo or model-free approximation, in which overall molecular tumbling and intramolecular motions are assumed to be uncorrelated and each characterized by a single exponential decay, the spectral density function becomes

\begin{equation}
\label{eqn:spin_relaxation_spectral_density}
J(\omega) = \left \langle \frac {1} {R_{AB}^6} \right \rangle \left( S^2 \frac {2 \tau_C} {1 + \omega^2 \tau_C^2} + \left( 1 - S^2 \right) \frac {2 \tau_{\mathsf{tot}}} {1 + \omega^2 \tau_{\mathsf{tot}}^2} \right)
\end{equation}

\noindent where $S^2$ (Eq. \ref{eqn:ls_order_parameter}) is the order parameter that characterizes the orientational freedom of the internuclear vector, $\tau_C$ is the time constant for overall molecular tumbling, $\tau_{\mathsf{int}}$ is the time constant for intramolecular motion, and $\tau_{\mathsf{tot}}$ is given by Eq. \ref{eqn:ls_total_tumbling}.

\section{The Solomon-Bloembergen equation for PRE}
\label{app:pre_solomon}

The relationship between structure, dynamics, and the transverse PRE $\Gamma_2$ is rigorously described by the Solomon-Bloembergen equation \cite{solomon_nuclear_1956}.

\begin{equation}
\label{eqn:solomon-bloembergen}
\Gamma_2 = \frac {K} {2} \left( 3 J(\omega_H) + 4 J(0) \right)
\end{equation}

\noindent where

\begin{equation}
\label{eqn:pre_constant}
K = \left( \frac {\mu_0} {4 \pi} \right)^2 \frac {\gamma_H^2 g^2 \mu_B^2} {10}
\end{equation}

\noindent and $\mu_0$ is the vacuum permeability, $g$ is the electron g-factor, $\mu_B$ is the magnetic moment of the free electron, $\gamma_H$ is the proton gyromagnetic ratio, $\omega_H$ is the Larmor frequency of the proton nucleus, and $J(\omega)$ is the spectral density function.
As introduced in Appendices \ref{app:noesy_distances} and \ref{app:spin_relax_redfield}, it is common to assume the Lipari-Szabo approximation in order to write the spectral density using an order parameter $S^2$.

\begin{equation}
\label{eqn:pre_spectral_density}
J(\omega) = \left \langle \frac {1} {R_{eH}^6} \right \rangle \left( S^2 \frac {2 \tau_C} {1 + \omega^2 \tau_C^2} + \left( 1 - S^2 \right) \frac {2 \tau_{\mathsf{tot}}} {1 + \omega^2 \tau_{\mathsf{tot}}^2} \right)
\end{equation}

where $R_{eH}$ is the distance between the amide proton and the paramagnetic electron.

\end{document}